\documentclass[journal]{IEEEtran}
\IEEEoverridecommandlockouts
\usepackage{cite}
\usepackage{amsmath,amssymb,amsfonts}
\usepackage{algorithm}
\usepackage{algorithmic}
\usepackage{graphicx}
\usepackage{textcomp}
\usepackage{xcolor}
\def\BibTeX{{\rm B\kern-.05em{\sc i\kern-.025em b}\kern-.08em
    T\kern-.1667em\lower.7ex\hbox{E}\kern-.125emX}}

\usepackage[utf8]{inputenc}
\usepackage{graphicx}
\usepackage{amsmath}
\usepackage{amsthm}
\usepackage{amssymb}
\usepackage{amsfonts}
\usepackage{comment}
\usepackage{cases,xspace,enumitem}
\usepackage{version}
\usepackage{stmaryrd}
\usepackage{subcaption}
\usepackage{hyperref}
\hypersetup{colorlinks=true, linkcolor=blue, breaklinks=true, urlcolor=blue}

\newtheorem{theorem}{Theorem}
\newtheorem{corollary}[theorem]{Corollary}
\newtheorem{lemma}[theorem]{Lemma}
\newtheorem{remark}[theorem]{Remark}
\newtheorem{prop}[theorem]{Proposition}

\newtheorem{assumption}{Assumption}
\newtheorem{example}{Example}

\usepackage[prependcaption,colorinlistoftodos]{todonotes}

\newcommand{\subscr}[2]{#1_{\textup{#2}}}

\newcommand{\setdef}[2]{\{#1 \; | \; #2\}}

\newcommand{\bigsetdef}[2]{\big\{#1 \; | \; #2\big\}}
\newcommand{\map}[3]{#1: #2 \rightarrow #3}

\newcommand{\integernonnegative}{\ensuremath{\mathbb{Z}}_{\ge 0}}
\newcommand{\real}{\ensuremath{\mathbb{R}}}

\newcommand{\realnonnegative}{\ensuremath{\mathbb{R}}_{\ge 0}}

\newcommand\oprocendsymbol{\hbox{$\triangle$}}
\newcommand\oprocend{\relax\ifmmode\else\unskip\hfill\fi\oprocendsymbol}

\renewcommand{\theenumi}{(\roman{enumi})}

\DeclareSymbolFont{bbold}{U}{bbold}{m}{n}
\DeclareSymbolFontAlphabet{\mathbbold}{bbold}
\newcommand{\vect}[1]{\mathbbold{#1}}
\newcommand{\vectorones}[1][]{\vect{1}_{#1}}

\newcommand{\Prob}{\mathbb{P}}
\newcommand{\E}{\mathbb{E}}

\newcommand{\seminorm}[1]{{\left\vert\kern-0.25ex\left\vert\kern-0.25ex\left\vert #1
        \right\vert\kern-0.25ex\right\vert\kern-0.25ex\right\vert}}

\newcommand{\semimeasure}[1]{\mu_{\seminorm{\cdot}}\kern-0.5ex\left(#1\right)}

\newcommand{\floor}[1]{\left\lfloor #1 \right\rfloor}

\newcommand{\mcN}{\mathcal{N}}

\newcommand{\mcW}{\mathcal{W}}

\newcommand{\mcZ}{\mathcal{Z}}

\newcommand{\FBess}[2]{\textrm{F}_{\textrm{BESS}}(#1, #2)}
\newcommand{\sstar}{s^{\star}}
\newcommand{\clip}[3]{\mathrm{clip}\left(#1\right)_{#2}^{#3}}

\newcommand{\bidup}[1]{b_{#1}^+}
\newcommand{\biddown}[1]{b_{#1}^-}
\newcommand{\bvec}[1]{\mathbf{b}_{#1}}      
\newcommand{\vc}{\mathbf{c}}                  
\newcommand{\vx}{\mathbf{x}}                  
\newcommand{\vy}{\mathbf{y}}                  
\newcommand{\regup}[1]{r_{#1}^+}
\newcommand{\regdn}[1]{r_{#1}^-}

\newcommand{\Tbid}{T_{\mathrm{bid}}}
\newcommand{\Tbidup}{\Tbid^+}

\newcommand{\Tbp}{T_{\mathrm{bp}}}

\newcommand{\Tbidadv}{T_{\mathrm{bidadv}}}

\newcommand{\meanpriceup}{\bar{\pi}^{+}}

\newcommand{\meanpriceupoff}{\bar{\pi}^{+}_{\mathrm{perf}}}

\newcommand{\worstscoreup}{\underline{\sigma}^{+}}
\newcommand{\worstscoredown}{\underline{\sigma}^{-}}
\newcommand{\effpriceup}{\widetilde{\pi}^{+}}
\newcommand{\effpricedn}{\widetilde{\pi}^{-}}
\newcommand{\CVaR}[1]{\text{CVaR}_{#1}}

\newcommand{\Sdn}{\underline{S}_k}
\newcommand{\Sup}{\overline{S}_k}

\newcommand{\Rdn}{\rho^{\downarrow}}
\newcommand{\Rup}{\rho^{\uparrow}}

\newcommand{\Dbid}{D_{\mathrm{bid}}}
\newcommand{\Dop}{D_{\mathrm{op}}}

\title{Degradation-Aware Participation of Battery Storage in
Frequency Regulation Markets}
\author{\IEEEauthorblockN{Anand Gokhale$^{1, 2}$ and Siddharth Bhela$^1$}\\
\IEEEauthorblockA{$^1$ Siemens Foundational Technology, Princeton, NJ, USA}\\
\IEEEauthorblockA{$^2$ University of California Santa Barbara, Santa Barbara, CA, USA \\
anand\_gokhale@ucsb.edu, siddharth.bhela@siemens.com}}

\begin{document}
\maketitle

\begin{abstract}
As data centers make the load on the power grid fluctuate faster, frequency
regulation grows more valuable.
Battery energy storage systems (BESSs) are well suited to provide frequency
regulation, but providing this service speeds up aging, especially during
operation at extreme state of charge (SoC).
In the PJM interconnection, the grid operator supplies a regulation signal
that a participating BESS is expected to follow, within a capacity that it
must bid before the signal is known.
Accordingly, we decompose the degradation-aware frequency regulation problem
into a bidding layer and an operations layer.
For bidding, we propose an adaptive strategy, assuming perfect tracking in which 
the bid depends on the SoC expected at the start of the bid period by means of 
choosing basepoints.
For operation, the BESS further protects its battery by deliberately giving
up a perfect performance score, to keep its SoC away from the extremes.
On the PJM regulation signal and prices of 2023, adaptive bids lengthen the
battery life by $31\%$ against bidding full capacity, for $2.2\%$ of the
revenue. Giving up a perfect score in operation lengthens the battery life of
constant bids by $10\%$ to $27\%$, at a change in revenue of at most $0.2\%$.
\end{abstract}

\section{Introduction}

\paragraph{Motivation}
Data centers are reshaping the load on the power grid. Their loads, especially
those training AI models, can fluctuate within
seconds~\cite{EC-etal:25,KBK-SM-VA:25}. Fluctuations this fast leave supply
and demand out of balance, and the grid frequency deviates. To hold the
frequency at its nominal value, system operators pay resources to adjust their
power to follow a regulation signal, a service called frequency regulation
(FR). As these loads grow, FR grows more valuable. In PJM, regulation prices
in 2024 were nearly double those of 2023, and have risen
since~\cite{PJM:DM2}.

Battery energy storage systems (BESSs) respond quickly and accurately to the
regulation signal, and markets that pay for the accuracy of the response
reward this~\cite{BX-YS-DSK-BZ:18}. In PJM, most BESSs provide only
regulation, and they made up $26.7\%$ of the regulation fleet in
2024~\cite{PJM:ESR26}. Unlike other resources, however, a BESS ages with every
cycle, and how fast depends on how it is operated~\cite{BX-JZ-TZ-EL-DSK:18}:
cycles at extreme state of charge (SoC) wear it faster~\cite{ZL-XW-KY-FZ:24}.
A BESS in FR must therefore earn revenue while keeping its battery alive for
many years. In PJM, it has two levers: its bid, which it commits before the
regulation signal is known, and its operational basepoint, which it sets in
real time.

\paragraph{Related work}
Early work manages the SoC of a BESS in regulation by scheduling its bids
and basepoints a day ahead, from an hourly averaged regulation
signal~\cite{JSY-JYC-GHA-YJC-MHK-DJW:16}. The schedule keeps the SoC within
fixed bounds, but does not model battery aging. The aging cost enters the bid
in~\cite{GH-QC-CK-PP-QX:16}, but without optimizing real-time operation.

Closest to our work is~\cite{BX-YS-DSK-BZ:18,YS-BX-YT-DK-BZ:19}. There, the
BESS bids the largest capacity that meets the performance requirement. It
protects the battery in real time, by no longer following the signal once a
cycle grows too deep. In PJM, however, a resource earns no regulation credit
in any five-minute interval whose performance score falls below
$0.25$~\cite{PJM:M28}, and is disqualified from the market if its score stays
low~\cite{PJM:M12}. The signal is also assumed to be energy neutral, so the
bid does not depend on the SoC. Other work sizes the bid from historical
signals~\cite{SL-JQ-AS-KP-PV:19,HZ-ZH-EM-SJM-YS:18,TM-DH-MRA:25} or
robustly~\cite{SJ-SB-NX:24}. These bids protect the performance score rather
than the battery.

In contrast, we set each bid from the SoC expected at the start of its bid
period. We account for the uncertainty of the regulation signal through the
conditional value at risk (CVaR) of the extreme SoCs. Instead of fixed SoC
bounds, we weigh in the bid a degradation proxy that penalizes extreme SoCs.
In real time, we likewise trade performance score for degradation, but within
bounds on the score that keep the BESS qualified. The BESS thus protects its
battery first through the bid, and deviates from the signal only within these
bounds.

\paragraph{Contributions}

\begin{itemize}
    \item We model how a BESS takes part in the PJM regulation market, with
    separate up and down bids, a basepoint it sets in real time, a performance
    score, and a bid committed before the signal is known. We split the problem
    into a bidding layer and an operations layer
    (Sections~\ref{sec:prelims} and~\ref{sec:formulation}).
    \item In the bidding layer, we weigh expected revenue against battery
    degradation, which depends on how far the SoC strays from its target. We
    show that the resulting problem is convex
    (Proposition~\ref{prop:convex}), and that a bidding rule is optimal for
    some degradation model exactly when it satisfies a simple monotonicity
    condition (Theorem~\ref{thm:inverse}).
    \item In the operations layer, the BESS can give up part of its performance
    score to keep its SoC near the target, either by letting the basepoint
    exceed the headroom or by following less of the signal when it moves the SoC
    away. We bound how much score each costs, so the BESS stays qualified, and
    we give conditions under which each is optimal
    (Section~\ref{sec:operations}).
    \item On the PJM regulation signal and prices of 2023, the price assumed
    when bidding trades revenue for battery life. Giving up a little score in
    operation lengthens the battery life of constant bids by $10$ to $27\%$ for
    a change in revenue of at most $0.2\%$, and earns $1.8\%$ more than bidding
    alone at equal time at extreme SoC (Section~\ref{sec:numerics}).
\end{itemize}

\section{Problem Formulation}\label{sec:formulation}

\paragraph{Notation.}
For $u \in \real$, $[u]_+ = \max(u, 0)$ and $[u]_- = \max(-u, 0)$. For an
interval $I$, $\Pi_I(x)$ is the point of $I$ closest to $x$, and
$\clip{x}{a}{b} = \Pi_{[a,b]}(x)$. For a random variable $X$ and
$\alpha \in (0,1)$, $\CVaR{1-\alpha}(X)$ is the mean of $X$ over its largest
$\alpha$ fraction of outcomes. For $u \in \real$, $\floor{u}$ is the largest
integer at most $u$. For a set $C$, $\mathbf{I}_C$ is its indicator, zero on $C$
and $+\infty$ elsewhere. For an event $\mathcal{E}$, $\iota_{\mathcal{E}}$ is one on
$\mathcal{E}$ and zero elsewhere. For $m \in \mathbb{N}$, $\vectorones[m] \in \real^m$ is the
vector of ones. For $\map{f}{\real^m}{\real \cup \{+\infty\}}$,
$f^*(\vy) = \sup_{\vx} \langle \vy, \vx\rangle - f(\vx)$ is its convex
conjugate and $\partial f$ its subdifferential.

We consider a BESS that participates in the PJM frequency regulation market
while remaining degradation aware. The state of charge (SoC) complicates the problem in two ways. First, the SoC may hit its bounds, preventing tracking of the regulation signal in one direction. Second, deep cycling of SoC accelerates BESS degradation.
\subsection{Model of the BESS}
Consider a BESS with rated power $P$, energy capacity $E$, and charging and
discharging efficiencies $\eta_c, \eta_d \in (0,1]$. We adopt the generator
convention: $u > 0$ denotes discharging and $u < 0$ charging. Applying power
$u$ over an interval of length $t$ changes the SoC from $s$ to
\begin{align}
    s^+ = \FBess{s}{u} = s - \frac{t}{E}\left( \frac{[u]_+}{\eta_d}
        - \eta_c\, [u]_- \right).
    \label{eq:soc}
\end{align}
With SoC limits $[s_{\min}, s_{\max}]$, a power $z$ can be delivered over an
interval of length $t$ from SoC $s$ only if $z \in \mcZ(s,t)$, where
\begin{align}
    \mcZ(s, t) := [-P, P] \cap \Big[ &-\frac{E(s_{\max}-s)}{\eta_c\, t},
        \frac{\eta_d\, E(s-s_{\min})}{t} \Big].
    \label{eq:feasible}
\end{align}
The operator recenters the SoC toward a target $\sstar \in (s_{\min}, s_{\max})$,
away from both limits.

\subsection{Market rules}
\subsubsection{Control regulations}
\begin{figure}[htb]
    \centering
    \includegraphics[width=\columnwidth]{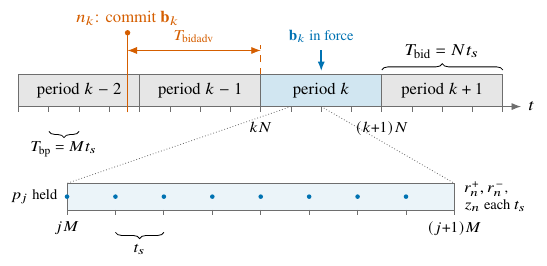}
    \caption{Timescales of the decisions. The bid $\bvec{k}$ is committed at
    sample $n_k$, $\Tbidadv$ before bid period $k$ begins, and is in force
    over the period. Each bid period contains several basepoint periods of
    length $\Tbp$ (top), over which the basepoint $p_j$ is held while the
    signal and the delivered power change every $t_s$ (bottom).}
    \label{fig:timeline}
\end{figure}

The BESS operator chooses its bids, its operational basepoint, and the power it
delivers, at three timescales, as shown in Fig.~\ref{fig:timeline}:
\begin{enumerate}
    \item Every $t_s$, the operator receives regulation signals
    $\regup{}, \regdn{} \in [0,1]$ for the up and down markets, and responds at
    the same rate. Samples are indexed by $n$.
    \item Every $\Tbp = Mt_s$, the operator sets its operational basepoint,
    which is held constant over the period. Basepoint periods are indexed by
    $j(n) = \floor{n / M}$, and the basepoint is denoted $p_j$, with the
    argument omitted.
    \item Each day is divided into bid periods of length
    $\Tbid = N t_s > \Tbp$, indexed by $k(n) = \floor{n/N}$. For each period,
    the operator commits fractions $\bidup{k}, \biddown{k} \in [0,1]$ of its
    capacity to the up and down markets, respectively, at least $\Tbidadv$
    before the period begins. We write
    $\bvec{k} = (\bidup{k}, \biddown{k}) \in [0,1]^2$ for the bid of period
    $k$. During $\Tbidadv$, the operator receives a cleared bid value from the RTO, which is bounded above by the bid.
\end{enumerate}
We assume the BESS is a price taker, and optimize the bids as if they clear in
full. Since the predicted SoC excursions are nondecreasing in the bid, as shown
in Section~\ref{sec:bidding}, partial clearing preserves the SoC limits.
The requested response at sample $n$ is
\begin{align}
    u_n = p_j + \bidup{k} P\, \regup{n-\delta} - \biddown{k} P\, \regdn{n-\delta},
    \label{eq:requested}
\end{align}
where $\delta \in \integernonnegative$ is a delay prescribed by the market. At
each sample, at most one of $\regup{n}$ and $\regdn{n}$ is nonzero. The
capacity not committed to regulation, $h_k^{\pm} = (1 - b_k^{\pm})P$, is the
\emph{headroom}; if $-h_k^{-} \le p_j \le h_k^{+}$, the request stays in
$[-P, P]$ for every signal, as shown in Fig.~\ref{fig:power}a.

\subsubsection{Revenue model}
Revenue is paid according to a performance score, which is defined only while
the corresponding market is active. Following the split-market redesign, PJM
uses a pure precision score: delivering $z_n$ when $u_n$ is requested yields
the scores
\begin{align}
    \sigma_n^{+} &=
    \begin{cases}
        \max\left(1 - \dfrac{|z_n - u_n|}{\bidup{k} P},\ 0\right) & \bidup{k} > 0,\\
        0 & \bidup{k} = 0,
    \end{cases}
    \label{eq:score_up} \\
    \sigma_n^{-} &=
    \begin{cases}
        \max\left(1 - \dfrac{|z_n - u_n|}{\biddown{k} P},\ 0\right) & \biddown{k} > 0,\\
        0 & \biddown{k} = 0.
    \end{cases}
    \label{eq:score_dn}
\end{align}
At each sample, the active market pays for performance and mileage at the
realized score, and the inactive market pays for performance at the historical
score:
\begin{align}
    R_n = \begin{cases}
        \bidup{k} P \sigma_n^{+}
          \big( \pi_{\mathrm{perf}, n}^{+} + \pi_{\mathrm{mil}, n}^{+}
          m_n^{+}/m_{\mathrm{ref}}^{+} \big) & \\
        \; + \biddown{k}P\, \bar\sigma_n^{-} \pi_{\mathrm{perf}, n}^{-},
          & \regup{n - \delta} > 0,\\[4pt]
        \biddown{k} P \sigma_n^{-}
          \big( \pi_{\mathrm{perf}, n}^{-} + \pi_{\mathrm{mil}, n}^{-}
          m_n^{-}/m_{\mathrm{ref}}^{-} \big) & \\
        \; + \bidup{k}P\, \bar\sigma_n^{+} \pi_{\mathrm{perf}, n}^{+},
          & \regdn{n - \delta} > 0,\\[4pt]
        \bidup{k}P\, \bar\sigma_n^{+} \pi_{\mathrm{perf}, n}^{+} & \\
        \; + \biddown{k}P\, \bar\sigma_n^{-} \pi_{\mathrm{perf}, n}^{-},
          & \text{otherwise,}
    \end{cases}
    \label{eq:revenue}
\end{align}
where $\pi_{\mathrm{perf}}^{\pm}$ are the performance prices,
$\pi_{\mathrm{mil}}^{\pm}$ the mileage prices,
$m_n^{+} = |\regup{n-\delta} - \regup{n-\delta-1}|$ and
$m_{n}^{-} = |\regdn{n-\delta} - \regdn{n-\delta-1}|$ the mileages,
$m_{\mathrm{ref}}^{\pm}$ reference historical mileages, and $\bar\sigma_n^{\pm}$
the historical performance scores defined in~\eqref{eq:histscore_def}. All
prices are nonnegative. The historical performance scores are trailing averages over the samples at
which each market is active. For a trailing window of length
$\subscr{T}{hist} = \subscr{H}{hist}\, t_s$, let
\begin{align}
    \mcW_n^{+} &= \bigsetdef{m}{n - \subscr{H}{hist} < m \le n,\ \regup{m-\delta} > 0},
    \nonumber\\
    \mcW_n^{-} &= \bigsetdef{m}{n - \subscr{H}{hist} < m \le n,\ \regdn{m-\delta} > 0},
\end{align}
and
\begin{align}
    \bar\sigma^{+}_n = \frac{1}{|\mcW_n^{+}|} \sum_{m \in \mcW_n^{+}} \sigma_m^{+},
    \qquad
    \bar\sigma^{-}_n = \frac{1}{|\mcW_n^{-}|} \sum_{m \in \mcW_n^{-}} \sigma_m^{-}.
    \label{eq:histscore_def}
\end{align}
A resource remains qualified only if
\begin{align}
    \bar\sigma^{+}_n \;\ge\; \subscr{\sigma}{min}
    \quad\text{and}\quad
    \bar\sigma^{-}_n \;\ge\; \subscr{\sigma}{min},
    \qquad \forall n.
    \label{eq:histscore}
\end{align}

\subsection{Optimization problem}
Consider a horizon $T = Ht_s$ with samples $\mcN = \{0,1,\ldots,H - 1\}$, and
let $D$ denote the degradation incurred over it, measured in units of revenue.
The regulation signal and the prices are random, so the operator chooses
policies: each bid, basepoint, and power may depend only on the information
available when it is chosen. Over the bids $\{\bvec{k}\}$,
basepoints $\{p_j\}$, and powers $\{z_n\}$, the operator solves
\begin{subequations}\label{eq:objective}
\begin{align}
    \max_{\bvec{}, p, z} \quad
    & \E\Big[\sum_{n\in\mcN} R_n - D\Big]
    \\
    \text{s.t.}\quad
    & s_{n+1}=\FBess{s_n}{z_n},\quad s_0 \text{ given}, \nonumber\\
    & z_n\in\mcZ(s_n,t_s), \nonumber\\
    & \bvec{k} \in [0,1]^2, \quad p_{j}\in [-P, P], \nonumber\\
    & \text{bids chosen $\Tbidadv$ ahead of the bid period},
      \label{eq:nonanticipative}\\
    & \eqref{eq:histscore}, \nonumber
\end{align}
\end{subequations}
for all $n \in \mcN$, with $k = k(n)$ and $j = j(n)$.

\subsection{Solution approach}
\begin{figure}[htb]
    \centering
    \includegraphics[width=0.9\columnwidth]{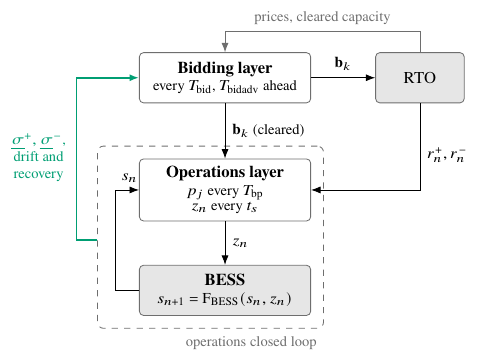}
    \caption{The two layers. The bidding layer commits $\bvec{k}$ to the
    RTO; the operations layer acts on the cleared capacity and the SoC. The
    bidding layer depends on the operations layer only through its
    worst-case scores (Assumption~\ref{ass:worst_score}) and the drift and
    recovery of the SoC (Assumption~\ref{ass:nonadversarial}).}
    \label{fig:architecture}
\end{figure}

Problem~\eqref{eq:objective} couples decisions at three timescales and is
intractable without forecasts. We decompose it into two layers
(Fig.~\ref{fig:architecture}):
\begin{enumerate}
    \item the bidding layer chooses $\bvec{k}$, $\Tbidadv$ before each bid
    period (Section~\ref{sec:bidding});
    \item the operations layer chooses the basepoint $p_j$ every $\Tbp$ and
    the power $z_n$ every $t_s$ (Section~\ref{sec:operations}).
\end{enumerate}
Each layer models the degradation $D$ differently: the bidding layer commits
before the SoC trajectory is known and penalizes the extreme SoCs of each bid
period by a function $\Dbid$ (Section~\ref{sec:bidding}), whereas the
operations layer observes the SoC and charges a function $\Dop$ of it at each
sample (Section~\ref{sec:operations}).

\section{Bidding Layer}\label{sec:bidding}
The bidding layer chooses the bids one period at a time, which limits the
forecast of the regulation signal to one bid period. Bid $k$ is committed at
sample $n_k$, $\Tbidadv$ before period $k$ begins. Let
$D_k$ be the degradation incurred in period $k$, and let
$\mcN_k = \{kN, \ldots, (k+1)N\}$ index the SoC over the period. Given the
information at $n_k$, the bid solves
\begin{align}
    \max_{\bvec{k} \in [0,1]^2} \quad
    & \E\Big[\sum_{n=kN}^{(k+1)N-1} R_n - D_k\Big]
    \label{eq:period_problem}\\
    \text{s.t.}\quad
    & \Prob\Big(\max_{n \in \mcN_k} s_n \ge s_{\max}\Big) \le \alpha,
    \nonumber\\
    & \Prob\Big(\min_{n \in \mcN_k} s_n \le s_{\min}\Big) \le \alpha,
    \nonumber
\end{align}
where the \emph{risk level} $\alpha \in (0,1)$ is the largest allowed
probability of reaching each SoC limit, and the expectation and
probabilities are over the regulation signal and the prices, given $s_{n_k}$
and the committed bids. Solving~\eqref{eq:period_problem} directly requires
the distribution of the SoC trajectory under the other layers' policies; we
instead make Assumptions~\ref{ass:worst_score} and~\ref{ass:avg_price} to bound the revenue, Assumption~\ref{ass:degradation} to estimate degradation, and Assumption~\ref{ass:nonadversarial} on the cooperation of the operations layer.


\subsection{Problem reduction}

The first two assumptions lower bound the expected revenue, which depends on
unknown prices and on the operations layers.
\begin{assumption}[Worst case score]\label{ass:worst_score}
    While the SoC remains within $(s_{\min}, s_{\max})$, the operations layer
    achieves scores of at least
    $\worstscoreup, \worstscoredown \in (\sigma_{\min}, 1]$ whenever the up and
    down market, respectively, is active.
\end{assumption}
As a consequence, the historical scores are also at least $\worstscoreup$ and
$\worstscoredown$. By~\eqref{eq:revenue}, at each sample the up bid earns
$\bidup{k} P$ times a score times the price
\begin{align}
    \pi_n^{+} =
    \begin{cases}
        \pi_{\mathrm{perf},n}^{+} + \pi_{\mathrm{mil},n}^{+}\,
            m_n^{+}/m_{\mathrm{ref}}^{+} & \regup{n-\delta} > 0,\\
        \pi_{\mathrm{perf},n}^{+} & \text{otherwise},
    \end{cases}
    \label{eq:sample_price}
\end{align}
and the down bid earns $\biddown{k} P$ times a score times $\pi_n^{-}$, defined
analogously.
\begin{assumption}[Average price]\label{ass:avg_price}
    The expected price per unit bid and unit score over a bid period is
    \begin{align}
        \E\Big[\sum_{n=kN}^{(k+1)N-1} \pi_n^{\pm}\Big] = \widetilde\pi^{\pm},
        \label{eq:effective_price}
    \end{align}
    and $\pi_n^{\pm} \le \pi_{\max}$ at every sample.
\end{assumption}
For instance, if the up market is active for a duration $\Tbidup$ in
expectation, with mean prices $\meanpriceup$ when active and $\meanpriceupoff$
when inactive, independent of which samples are active, then
$\effpriceup = \meanpriceup \Tbidup/t_s + (N - \Tbidup/t_s)\meanpriceupoff$.

\begin{lemma}[Revenue lower bound]\label{lem:revenue}
    Let Assumptions~\ref{ass:worst_score} and~\ref{ass:avg_price} hold. Then
    the expected revenue of bid period $k$ satisfies
    \begin{align}
        \E \Big[ \sum_{n = kN}^{(k+1)N - 1}R_n\Big] \;\geq\;
        &\bidup{k}\worstscoreup P \big(\effpriceup - \varepsilon_k N \pi_{\max}\big)
        \nonumber\\
        + &\biddown{k}\worstscoredown P \big(\effpricedn - \varepsilon_k N \pi_{\max}\big),
        \label{eq:revenue_bid}
    \end{align}
    where $\varepsilon_k$ is the probability that the SoC reaches $s_{\min}$ or
    $s_{\max}$ during bid period $k$.
\end{lemma}
\begin{proof}
    Let $\mathcal{E}_k$ be the event that the SoC remains in
    $(s_{\min}, s_{\max})$ over bid period $k$, so that
    $\varepsilon_k = \Prob(\mathcal{E}_k^{c})$, and let $\underline R_n$ denote
    $R_n$ with every score, realized or historical,
    replaced by $\worstscoreup$ or $\worstscoredown$. On $\mathcal{E}_k$,
    Assumption~\ref{ass:worst_score} and nonnegative prices give
    $R_n \ge \underline R_n$; off $\mathcal{E}_k$, $R_n \ge 0$. Hence
    \begin{align*}
        \E \sum_n R_n \ge \E\Big[\sum_n \underline R_n\, \iota_{\mathcal{E}_k}\Big]
        = \E \sum_n \underline R_n - \E\Big[\sum_n \underline R_n\,
            \iota_{\mathcal{E}_k^c}\Big].
    \end{align*}
    By~\eqref{eq:sample_price}, $\underline R_n = \bidup{k}\worstscoreup P \pi_n^{+}
    + \biddown{k}\worstscoredown P \pi_n^{-}$. By
    Assumption~\ref{ass:avg_price}, the first expectation equals
    $\bidup{k}\worstscoreup P \effpriceup + \biddown{k}\worstscoredown P \effpricedn$,
    and since $\pi_n^{\pm} \le \pi_{\max}$, the second is at most
    $(\bidup{k}\worstscoreup + \biddown{k}\worstscoredown) P N \pi_{\max} \varepsilon_k$.
\end{proof}
Degradation models such as rainflow counting use the whole SoC trajectory,
which is unknown when the bids are committed. We therefore model the
degradation of a bid period by its extreme SoCs.

\begin{assumption}[Degradation]\label{ass:degradation}
    Let $\mathcal{T} = \setdef{(s_1, s_2)}{0 \le s_1 \le s_2 \le 1}$, and let
    $\map{\Dbid}{\mathcal{T}}{\realnonnegative}$ be convex, lower semicontinuous,
    nonincreasing in its first argument, and nondecreasing in its second. The
    degradation incurred
    in bid period $k$, in units of revenue, is $D_k = \Dbid(s_1, s_2)$, where
    $s_1 = \min_{n \in \mcN_k} s_n$ and $s_2 = \max_{n \in \mcN_k} s_n$.
\end{assumption}
Thus $\Dbid$ can penalize the cycle depth, through $s_2 - s_1$, and the
extreme SoCs, through $\sstar - s_1$ and $s_2 - \sstar$. Estimating $s_1$ and
$s_2$ requires one more assumption on the operations layer.

\begin{figure}[htb]
    \centering
    \includegraphics[width=\columnwidth]{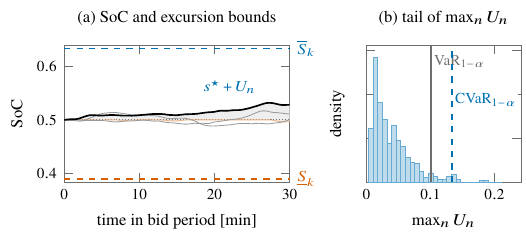}
    \caption{Predicted extremes over one 30-minute bid period of RegD, symmetric
    bid $0.99$, $\alpha = 0.05$. (a) SoC under~\eqref{eq:recovered} for three regulation
    signals, the bounds $\sstar + U_n$ and $\sstar - L_n$ of the bold
    trajectory, and the extremes~\eqref{eq:extremes}. (b) Distribution of
    $\max_n U_n$ over $400$ bid periods sampled from the training days, with
    its value at risk
    and conditional value at risk at level $1-\alpha$.}
    \label{fig:excursions}
\end{figure}

\begin{assumption}[Non-adversarial operations]\label{ass:nonadversarial}
    While the SoC is strictly between its limits, the basepoint uses at least
    the headroom to move the SoC toward $\sstar$, and the delivered power
    departs from the request only to move the SoC toward $\sstar$.
\end{assumption}
Under Assumption~\ref{ass:nonadversarial}, tracking the regulation signal
moves the SoC at sample $n$ by the \emph{drift}
\begin{align}
    \Delta_n = \frac{t_s P}{E}
        \Big(\eta_c\, \biddown{k(n)}\, \regdn{n-\delta}
            - \frac{\bidup{k(n)}}{\eta_d}\, \regup{n-\delta}\Big),
    \label{eq:drift}
\end{align}
and the basepoint then moves it back toward $\sstar$ by at least the
\emph{recovery}
\begin{align}
    \Rdn_n = \frac{t_s h^{+}_{k(n)}}{\eta_d E},
    \qquad
    \Rup_n = \frac{\eta_c t_s h^{-}_{k(n)}}{E},
    \label{eq:recovery}
\end{align}
from above and below $\sstar$, respectively, unless it reaches $\sstar$.
Applying the recovery at every sample, without passing $\sstar$, gives
\begin{align}
    s_{n+1} = \tilde s_n + \clip{\sstar - \tilde s_n}{-\Rdn_n}{\Rup_n},
    \qquad \tilde s_n = s_n + \Delta_n.
    \label{eq:recovered}
\end{align}
\begin{remark}[Sample level approximation]\label{rem:overshoot}
    The basepoint is updated only every $\Tbp$, so it may pass $\sstar$.
    Applying the recovery at every sample simplifies the analysis, at a
    worst-case error of order $\Tbp/(E/P)$.
\end{remark}
The extremes are predicted from~\eqref{eq:recovered} in two steps. First, the
bids in force between $n_k$ and $kN$ are already committed, so the SoC at the start
of the bidding period is estimated by
\begin{align}
    \hat s_k = \E\big[s_{kN}\big],
    \label{eq:entry_estimate}
\end{align}
with $s_n$ evolving from $s_{n_k}$ by the drift~\eqref{eq:drift} and the full
recovery~\eqref{eq:recovery} toward $\sstar$, neglecting the stop at $\sstar$
in~\eqref{eq:recovered}. Since $s_{kN}$ is then $s_{n_k}$ plus the drifts and
recoveries over the lead time, $\hat s_k$ is affine in $s_{n_k}$ and the bids
in force. Like the extremes below, we
fit it once, offline, on historical signals. Neglecting the stop misestimates $\hat s_k$ by
at most the recovery~\eqref{eq:recovery} summed over the lead time, which is
small for bids near one. Second, over the
bid period, the SoC stays within the upper and lower \emph{excursions} $U_n$
and $L_n$ from $\sstar$, $\sstar - L_n \le s_n \le \sstar + U_n$, where
\begin{align}
    U_{n+1} &= \big[U_n + \Delta_n - \Rdn_n\big]_+,
    & U_{kN} &= [\hat s_k - \sstar]_+,
    \nonumber\\
    L_{n+1} &= \big[L_n - \Delta_n - \Rup_n\big]_+,
    & L_{kN} &= [\sstar - \hat s_k]_+,
    \label{eq:excursions}
\end{align}
by induction on~\eqref{eq:recovered}. Up to the error of
Remark~\ref{rem:overshoot}, \eqref{eq:recovered} is the slowest return to
$\sstar$ that Assumption~\ref{ass:nonadversarial} allows, so these bounds
hold for any operations layer that satisfies it.
At the risk level $\alpha$, the predicted extremes are
\begin{subequations}\label{eq:extremes}
\begin{align}
    \Sdn &= \sstar - \CVaR{1-\alpha}\Big(\max_{n \in \mcN_k} L_n\Big),
    \label{eq:extremes_lo}\\
    \Sup &= \sstar + \CVaR{1-\alpha}\Big(\max_{n \in \mcN_k} U_n\Big).
    \label{eq:extremes_hi}
\end{align}
\end{subequations}
The bounds~\eqref{eq:excursions} and the extremes~\eqref{eq:extremes} are
illustrated in Fig.~\ref{fig:excursions}.

Let the \emph{price vector} be
$\vc = P(\worstscoreup\effpriceup, \worstscoredown\effpricedn)$, so that the
leading term of~\eqref{eq:revenue_bid} is $\langle \vc, \bvec{k}\rangle$.
Replacing in~\eqref{eq:period_problem} the revenue by
$\langle \vc, \bvec{k}\rangle$, the degradation by $\Dbid$ at the predicted
extremes, and the SoC limits by limits on these predictions gives
\begin{align}
    \max_{\bvec{k}} \quad
    & \langle \vc, \bvec{k}\rangle
      - \Dbid\big(\Sdn(\bvec{k}),\, \Sup(\bvec{k})\big)
    \label{eq:simplified}\\
    \text{s.t.}\quad
    & \Sdn(\bvec{k}) \ge s_{\min}, \quad \Sup(\bvec{k}) \le s_{\max},
      \quad \bvec{k} \in [0,1]^2,
      \nonumber
\end{align}
Problem~\eqref{eq:simplified} depends only on $\vc$, $\hat s_k$, and the
distribution of the regulation signal, all known at $n_k$, so its solution
satisfies~\eqref{eq:nonanticipative}.
The omitted term of~\eqref{eq:revenue_bid} is small, since
$\varepsilon_k \le 2\alpha$ under the constraints of~\eqref{eq:simplified}, by
Corollary~\ref{cor:limits}.
We now study the convexity of problem~\eqref{eq:simplified}.

\begin{prop}[Convexity of the bid problem]\label{prop:convex}
    Let Assumption~\ref{ass:degradation} hold. Then $\Sdn$ is concave and
    $\Sup$ is convex in $\bvec{k}$, and problem~\eqref{eq:simplified} is a
    convex optimization problem. It is feasible whenever
    $\hat s_k \in [s_{\min}, s_{\max}]$.
\end{prop}
\begin{proof}
    Over the bid period, $\Delta_n - \Rdn_n$ is affine in $\bvec{k}$ for each
    realization of the signal. Unrolling~\eqref{eq:excursions},
    \begin{align*}
        U_n = \max\Big\{&U_{kN} + \sum_{i=kN}^{n-1}(\Delta_i - \Rdn_i),\\
            &\max_{kN < m \le n} \sum_{i=m}^{n-1}(\Delta_i - \Rdn_i)\Big\},
    \end{align*}
    with the empty sum equal to zero, is a maximum of affine functions, and so
    is $\max_{n \in \mcN_k} U_n$; both are convex in $\bvec{k}$. Since
    $\CVaR{1-\alpha}$ is convex and monotone~\cite{PA-FD-JME-DH:99}, $\Sup$ is
    convex; likewise, $\Sdn$ is concave. As $\Dbid$ is convex, nonincreasing in its
    first argument, and nondecreasing in its second, its composition with
    $(\Sdn, \Sup)$ is convex, so the objective of~\eqref{eq:simplified} is
    concave. The
    constraint sets are a superlevel set of a concave function, a sublevel set
    of a convex function, and $[0,1]^2$, all convex. At $\bvec{k} = 0$,
    $(\Sdn, \Sup) = (\min\{\hat s_k, \sstar\}, \max\{\hat s_k, \sstar\})$, which
    is feasible whenever $\hat s_k \in [s_{\min}, s_{\max}]$.
\end{proof}
Next, we discuss some properties of the predicted extremes.
\begin{lemma}[Absorbed drift]\label{lem:absorbed}
    Over bid period $k$:
    \begin{enumerate}
        \item\label{it:absorbed_up} if $\eta_c\eta_d\,\biddown{k} \le 1 - \bidup{k}$, then $U_n$ is
        nonincreasing in $n$;
        \item\label{it:absorbed_dn} if $\bidup{k} \le \eta_c\eta_d\,(1 - \biddown{k})$, then $L_n$ is
        nonincreasing in $n$;
        \item if the conditions of items~\ref{it:absorbed_up} and~\ref{it:absorbed_dn} hold, then $\Sdn = \min\{\hat s_k, \sstar\}$ and
        $\Sup = \max\{\hat s_k, \sstar\}$.
    \end{enumerate}
\end{lemma}
\begin{proof}
    At each sample, at most one market is active. If the up market is active,
    $\Delta_n \le 0 \le \Rdn_n$. If the down market is active,
    by~\eqref{eq:drift} and~\eqref{eq:recovery},
    $\Delta_n \le \eta_c t_s P \biddown{k}/E
    \le t_s P (1 - \bidup{k})/(\eta_d E) = \Rdn_n$. In both cases
    $U_{n+1} = [U_n + \Delta_n - \Rdn_n]_+ \le U_n$. Symmetrically,
    $-\Delta_n \le t_s P \bidup{k}/(\eta_d E) \le \eta_c t_s P(1 - \biddown{k})/E
    = \Rup_n$, so $L_{n+1} \le L_n$. The maxima of $U_n$ and $L_n$ over the
    period are then their initial values in~\eqref{eq:excursions}, which do
    not depend on the signal, and~\eqref{eq:extremes} gives the extremes.
\end{proof}
\begin{remark}[Properties of the predicted extremes]\label{rem:properties}
\begin{enumerate}
    \item\label{it:monotone} Larger bids widen the extremes, so when we clear less than we bid, we preserve
    feasibility of~\eqref{eq:simplified}.
    \item Without losses, Lemma~\ref{lem:absorbed} simplifies to
    $\bidup{k} + \biddown{k} \le 1$. Motivated by this finding, we restrict the 
    bids to $[1/2, 1]^2$ for the rest of this work.
    \item\label{it:headroom} At a full bid, there is no headroom, so the
    recovery~\eqref{eq:recovery} is zero and the SoC follows the
    drift~\eqref{eq:drift} of the regulation signal alone.
\end{enumerate}
\end{remark}
\begin{corollary}[SoC limits]\label{cor:limits}
    Let Assumption~\ref{ass:nonadversarial} hold, let the SoC start the bid
    period at $\hat s_k$, and let $\max_{n \in \mcN_k} U_n \neq s_{\max} - \sstar$
    and $\max_{n \in \mcN_k} L_n \neq \sstar - s_{\min}$ almost surely. Then
    every bid feasible for~\eqref{eq:simplified} satisfies the chance
    constraints of~\eqref{eq:period_problem}, so $\varepsilon_k \le 2\alpha$.
\end{corollary}
\begin{proof}
    Let $X = \max_{n \in \mcN_k} U_n$. The $(1-\alpha)$-quantile $q$ of $X$
    satisfies $\Prob(X > q) \le \alpha$, and
    $\CVaR{1-\alpha}(X) \ge q$~\cite{RTR-SU:00}. Since
    $s_n \le \sstar + U_n$ and $X \neq s_{\max} - \sstar$ almost surely, the constraint
    $\Sup \le s_{\max}$ gives
    $\Prob(\max_{n \in \mcN_k} s_n \ge s_{\max}) \le \Prob(X \ge s_{\max} - \sstar)
    = \Prob(X > s_{\max} - \sstar) \le \Prob(X > q) \le \alpha$, which is the
    first chance constraint; the
    same argument applied to $L_n$ and $\Sdn$ gives the second. A union bound then gives
    $\varepsilon_k \le 2\alpha$.
\end{proof}
We estimate the CVaRs from $K$ historical bid periods of the regulation
signal, using the Rockafellar--Uryasev representation~\cite{RTR-SU:00}.
Problem~\eqref{eq:simplified} is then a convex program with $O(KN)$ variables,
one per sample of each bid period. It is a linear program when $\Dbid$ is
piecewise linear. For one year of history, $KN$ is about $1.6 \times 10^7$.
We therefore fit, once and offline, an affine model of the predicted extremes for
bids in $[1/2, 1]^2$, validated in Section~\ref{sec:fit}:
\begin{subequations}\label{eq:affine_split}
\begin{align}
    \Sdn &\approx \hat x_k^{\downarrow}
        - \bidup{k}\, g^{+}_{\downarrow} - \biddown{k}\, g^{-}_{\downarrow},
        \label{eq:affine_lo}\\
    \Sup &\approx \hat x_k^{\uparrow}
        + \bidup{k}\, g^{+}_{\uparrow} + \biddown{k}\, g^{-}_{\uparrow},
        \label{eq:affine_hi}
\end{align}
\end{subequations}
where the intercepts $\hat x_k^{\downarrow} = \min\{\hat s_k, \sstar\} - g^{0}_{\downarrow}$
and $\hat x_k^{\uparrow} = \max\{\hat s_k, \sstar\} + g^{0}_{\uparrow}$ do not
depend on the bid. In matrix form,
\begin{align}
    \big(\Sdn, \Sup\big) \approx \hat\vx_k + A\, \bvec{k},
    \qquad
    A = \begin{bmatrix}
        -g^{+}_{\downarrow} & -g^{-}_{\downarrow}\\
        \phantom{-}g^{+}_{\uparrow} & \phantom{-}g^{-}_{\uparrow}
    \end{bmatrix},
    \label{eq:affine_map}
\end{align}
with $\hat\vx_k = (\hat x_k^{\downarrow}, \hat x_k^{\uparrow})$.

\begin{algorithm}[t]
\caption{Bidding for bid period $k$}\label{alg:bidding}
\begin{algorithmic}[1]
\REQUIRE SoC $s_{n_k}$ at commitment; bids in force until period $k$ begins;
    fitted $\hat s_k$ and affine extremes~\eqref{eq:affine_split};
    $\underline\sigma^{\pm}$, $\widetilde\pi^{\pm}$, $\Dbid$
\STATE Evaluate the fitted $\hat s_k$~\eqref{eq:entry_estimate} at $s_{n_k}$
    and the bids in force
\STATE Evaluate the intercepts $\hat\vx_k$ of~\eqref{eq:affine_split} at $\hat s_k$
\STATE Solve the convex problem~\eqref{eq:simplified} with the affine
    extremes~\eqref{eq:affine_split} over bids in $[1/2, 1]^2$
\RETURN $\bvec{k}$
\end{algorithmic}
\end{algorithm}

\subsection{Closed-form analysis}
Taking~\eqref{eq:affine_map} as exact, we derive the optimal bids in closed
form: first for a general $\Dbid$. We extend $\Dbid$ by $+\infty$
outside its domain $\mathcal{T}$ (from Assumption~\ref{ass:degradation}), so that it is a closed convex function on $\real^2$, and
let
\begin{align*}
    \mathcal{X}_k = \big\{\hat\vx_k + A \bvec{} : \bvec{} \in [1/2,1]^2\big\}
        \cap \big([s_{\min}, 1] \times [0, s_{\max}]\big)
\end{align*}
be the set of extremes attainable by a feasible bid in $[1/2, 1]^2$.

\begin{prop}[Closed-form bid]\label{prop:closed_form}
    Let Assumption~\ref{ass:degradation} hold, let $\mathcal{X}_k$ be
    nonempty, and let $A$ be invertible. Define
    $\vy = (A^{\top})^{-1} \vc$. Then $\bvec{k}^\star$
    solves~\eqref{eq:simplified} over bids in $[1/2, 1]^2$ if and only if
    \begin{align}
        \bvec{k}^\star = A^{-1}\big(\vx^\star - \hat\vx_k\big),
        \qquad
        \vx^\star \in \partial\big(\Dbid + \mathbf{I}_{\mathcal{X}_k}\big)^{*}(\vy),
        \label{eq:closed_form_split}
    \end{align}
    Moreover, if some
    $\tilde{\vx} \in \partial \Dbid^{*}(\vy)$ lies in $\mathcal{X}_k$, then
    $\vx^\star = \tilde{\vx}$ is optimal.
\end{prop}
\begin{proof}
    Since $A$ is invertible, $\vx = \hat\vx_k + A \bvec{}$ maps the
    bids in $[1/2, 1]^2$ feasible for~\eqref{eq:simplified} bijectively onto
    $\mathcal{X}_k$, and
    the revenue becomes
    $\langle \vc, A^{-1}(\vx - \hat\vx_k)\rangle
    = \langle \vy, \vx\rangle - \langle \vy, \hat\vx_k\rangle$.
    Hence~\eqref{eq:simplified} maximizes
    $\langle \vy, \vx\rangle - (\Dbid + \mathbf{I}_{\mathcal{X}_k})(\vx)$. By
    Assumption~\ref{ass:degradation}, $\Dbid$ is closed and convex;
    $\mathcal{X}_k$ is compact, convex, and nonempty. Hence
    $\Dbid + \mathbf{I}_{\mathcal{X}_k}$ is closed, proper, and convex, and its
    conjugate is finite everywhere. By the Fenchel--Young
    equality~\cite[Thm.~23.5]{RTR:70}, $\vx$ maximizes
    $\langle \vy, \vx\rangle - (\Dbid + \mathbf{I}_{\mathcal{X}_k})(\vx)$ if and only if
    $\vx \in \partial (\Dbid + \mathbf{I}_{\mathcal{X}_k})^*(\vy)$, which is nonempty. Finally, $\tilde{\vx} \in \partial \Dbid^*(\vy)$
    maximizes $\langle \vy, \vx\rangle - \Dbid(\vx)$ over all $\vx$, hence
    over $\mathcal{X}_k$ when $\tilde{\vx} \in \mathcal{X}_k$.
\end{proof}
\begin{remark}[Conditioning of $A$]\label{rem:conditioning}
By Remark~\ref{rem:properties}\ref{it:monotone}, both columns of $A$ lie in the same quadrant.
If they are nearly parallel, $A$ is poorly conditioned, and $A^{-1}$ amplifies
errors in the extremes. 
\end{remark}

To avoid inverting $A$, we study symmetric bids, and split bids with one bid
held at one. For symmetric
bids $\bvec{k} = \beta_k \vectorones[2]$,
problem~\eqref{eq:simplified} becomes
\begin{align}
    \max_{\beta_k \in [1/2, 1]} \quad
    & \beta_k \langle \vc, \vectorones[2]\rangle
      - \Dbid\big(\hat\vx_k + \beta_k A\vectorones[2]\big)
    \label{eq:symmetric_problem}\\
    \text{s.t.}\quad
    & \hat\vx_k + \beta_k A\vectorones[2] \in [s_{\min}, 1] \times [0, s_{\max}].
    \nonumber
\end{align}

\begin{corollary}[Symmetric bid]\label{cor:symmetric}
    Let Assumption~\ref{ass:degradation} hold. Then the solution
    of~\eqref{eq:symmetric_problem} is the smallest feasible $\beta_k$ at
    which the right derivative of $\beta \mapsto \Dbid(\hat\vx_k + \beta A\vectorones[2])$
    is at least $\langle \vc, \vectorones[2]\rangle$, and the largest feasible
    $\beta_k$ if there is none.
\end{corollary}
\begin{proof}
    The feasible set of~\eqref{eq:symmetric_problem} is an interval. Since
    $\Dbid$ is convex and its argument is affine in $\beta_k$, the objective
    is concave, and its right derivative is nonincreasing. The objective
    therefore increases up to the first $\beta_k$ at which this derivative is
    nonpositive and does not increase after it.
\end{proof}
The right derivative of $\beta \mapsto \Dbid(\hat\vx_k + \beta A\vectorones[2])$ is
nondecreasing, so the bid may be found by bisection.

\begin{figure}[htb]
    \centering
    \includegraphics[width=\columnwidth]{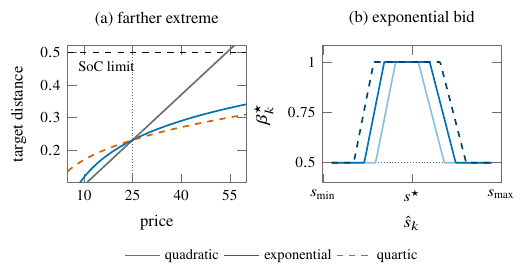}
    \caption{Penalty shapes of Example~\ref{ex:shapes}, with the affine
    extremes~\eqref{eq:affine_split} and, for each shape, $\lambda$ chosen so that at a
    price of $25$\,\$/MW the bid is full for $\hat s_k \in [0.4, 0.6]$. (a) Distance
    of the farther extreme from $\sstar$ at the optimal bid, against the price.
    (b) Optimal symmetric bid~\eqref{eq:exp_bid} under the exponential shape
    against the predicted starting SoC, at prices $15$, $25$, and $40$ (light
    to dark).}
    \label{fig:penalty}
\end{figure}

We illustrate Corollary~\ref{cor:symmetric} with an example.
\begin{example}[Penalty shapes]\label{ex:shapes}
Let the degradation of a bid period grow with the distance from $\sstar$ of
its extreme SoC farther from it,
\begin{align}
    \Dbid(s_1, s_2) = \lambda\, \psi\big(\max\{\sstar - s_1,\ s_2 - \sstar\}\big),
    \label{eq:penalty}
\end{align}
with $\lambda > 0$ and $\psi(d) = (2d)^2$, $(2d)^4$, or
$(\exp(8d) - 1)/(\exp(4) - 1)$ for $d \ge 0$, and $\psi(d) = 0$ for $d < 0$.
Each $\psi$ is convex and nondecreasing, so $\Dbid$ satisfies
Assumption~\ref{ass:degradation}. Let $\hat s_k$ be high enough that the upper
extreme is the farther one, at distance $d$ from $\sstar$.

If the optimal bid lies strictly between its bounds, then, by
Corollary~\ref{cor:symmetric},
\begin{align}
    \lambda (A\vectorones[2])_2\, \psi'(d) = \langle \vc, \vectorones[2]\rangle,
    \label{eq:interior_distance}
\end{align}
so $d$ grows linearly with the price for the quadratic shape, but only with
its cube root and its logarithm for the quartic and exponential shapes, as
shown in Fig.~\ref{fig:penalty}a.
\end{example}

Another way to avoid inverting $A$ is to hold one bid at one and choose the
other. When $\hat s_k < \sstar$, the down signal moves the SoC back toward
$\sstar$, so we hold $\biddown{k} = 1$ and choose $\bidup{k}$ by
Corollary~\ref{cor:symmetric}, with $A(\bidup{k}, 1)$ in place of
$\beta_k A\vectorones[2]$ and $P\worstscoreup\effpriceup$ in place of
$\langle \vc, \vectorones[2]\rangle$; when $\hat s_k \ge \sstar$, we hold
$\bidup{k} = 1$ and choose $\biddown{k}$ likewise. 

\subsection{Inverse problem: the penalty implied by a bidding rule}
Conversely, operators often bid by heuristic rules that map prices to bids.
At a fixed starting SoC, a rule $\map{\gamma}{\mathcal{D}}{[0,1]^2}$ on a set
$\mathcal{D} \subseteq \real^2$ of price vectors is \emph{cyclically monotone}
if, for all $\vc_0, \ldots, \vc_m \in \mathcal{D}$ and $\vc_{m+1} = \vc_0$,
\begin{align}
    \sum_{i=0}^{m} \big\langle \vc_{i+1} - \vc_i,\ \gamma(\vc_i) \big\rangle \le 0.
    \label{eq:cyclic}
\end{align}

\begin{theorem}[Rationalizable bidding rules]\label{thm:inverse}
    A bidding rule $\map{\gamma}{\mathcal{D}}{[0,1]^2}$ is cyclically monotone if
    and only if there is a closed, proper, convex $\varphi$ with
    $\operatorname{dom} \varphi \subseteq [0,1]^2$ such that, for every
    $\vc \in \mathcal{D}$, $\gamma(\vc)$ maximizes
    $\langle \vc, \bvec{}\rangle - \varphi(\bvec{})$. One such $\varphi$ is
    $\theta^* + \mathbf{I}_{[0,1]^2}$, where, for any fixed $\bar{\vc} \in \mathcal{D}$,
    \begin{align}
        \theta(\vc) = \sup \sum_{i=0}^{m} \big\langle \vc_{i+1} - \vc_i,\ \gamma(\vc_i) \big\rangle,
        \label{eq:rockafellar}
    \end{align}
    over all $m$ and $\vc_1, \ldots, \vc_m \in \mathcal{D}$, with
    $\vc_0 = \bar{\vc}$ and $\vc_{m+1} = \vc$. If, in addition, $A$ is
    invertible and each $\gamma(\vc)$ satisfies the SoC limits
    of~\eqref{eq:simplified}, then $\gamma(\vc)$ is optimal for~\eqref{eq:simplified}
    with $\Dbid(\vx) = \varphi(A^{-1}(\vx - \hat\vx_k))$.
\end{theorem}
\begin{proof}
    If $\varphi$ exists, then $\gamma(\vc) \in \partial \varphi^*(\vc)$ by the Fenchel--Young
    equality~\cite[Thm.~23.5]{RTR:70}, and summing the subgradient
    inequalities
    $\varphi^*(\vc_{i+1}) \ge \varphi^*(\vc_i) + \langle \vc_{i+1} - \vc_i, \gamma(\vc_i)\rangle$
    around the cycle gives~\eqref{eq:cyclic}. Conversely, if $\gamma$ is cyclically
    monotone, then $\theta$ is closed, proper, and convex with
    $\gamma(\vc) \in \partial \theta(\vc)$~\cite{RTR:66},~\cite[Thm.~24.8]{RTR:70}, so
    $\gamma(\vc)$ maximizes $\langle \vc, \bvec{}\rangle - \theta^*(\bvec{})$ over all
    $\bvec{}$, and hence over $[0,1]^2$. The last claim follows from the change
    of variables in the proof of Proposition~\ref{prop:closed_form}.
\end{proof}

\begin{corollary}[Symmetric bidding rules]\label{cor:inverse_symmetric}
    A symmetric rule $\map{\gamma}{\mathcal{D}}{[0,1]}$, on an interval
    $\mathcal{D}$ of revenues $q$ per unit bid, maximizes
    $q\beta - \varphi(\beta)$ for every $q \in \mathcal{D}$ and some closed,
    proper, convex $\varphi$ with $\operatorname{dom}\varphi \subseteq [0,1]$
    if and only if $\gamma$ is nondecreasing.
\end{corollary}
\begin{proof}
    On the real line, cyclic monotonicity is monotonicity; the claim follows
    from Theorem~\ref{thm:inverse}.
\end{proof}

\begin{figure}[htb]
    \centering
    \includegraphics[width=\columnwidth]{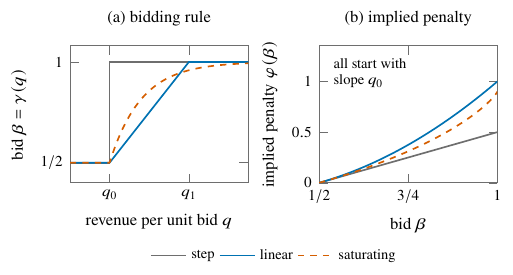}
    \caption{(a) Three symmetric bidding rules with bids in $[1/2, 1]$. (b) The
    penalties implied by Corollary~\ref{cor:inverse_symmetric}, normalized to
    $\varphi(1/2) = 0$: the step rule implies a linear penalty, the linear rule
    a quadratic one, and the saturating rule one between the two.}
    \label{fig:implied_penalty}
\end{figure}

\begin{remark}[Interpreting a heuristic]\label{rem:inverse}
For a split-bid rule that is smooth in the prices, Theorem~\ref{thm:inverse}
gives a simple test: raising the down price must change the up bid exactly as
much as raising the up price changes the down bid. The implied penalty is
known only on the bids the rule uses.
\end{remark}
For a nondecreasing $\gamma$, one such $\varphi$ is
$\theta^* + \mathbf{I}_{[0,1]}$, with
$\theta(q) = \int_{\bar q}^{q} \gamma(\tau)\, d\tau$ for any
$\bar q \in \mathcal{D}$, since $\theta$ is convex with
$\gamma(q) \in \partial\theta(q)$. We apply this
to three rules that bid $1/2$ below a revenue $q_0$ per unit bid, as shown in
Fig.~\ref{fig:implied_penalty}. Each implied
penalty has slope $q_0$ at a bid of $1/2$: a step rule implies the linear
penalty $q_0(\beta - 1/2)$, a rule rising linearly up to $q_1$ adds the
quadratic term $(q_1 - q_0)(\beta - 1/2)^2$. For the parameters of
Fig.~\ref{fig:implied_penalty}, a saturating rule implies a penalty between
these two, but whose marginal cost grows without bound as $\beta \to 1$.

\section{Operations Layer}\label{sec:operations}
\begin{figure}[htb]
    \centering
    \includegraphics[width=0.85\columnwidth]{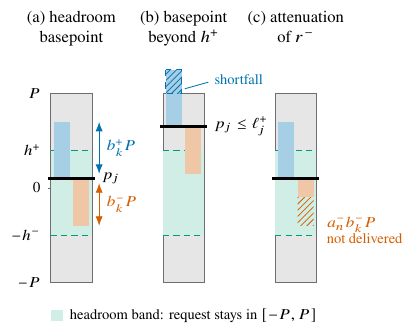}
    \caption{Range of the requested power about the basepoint $p_j$, with the
    up (blue) and down (orange) components. (a) Basepoint within the
    headroom: the request stays in $[-P, P]$. (b) Basepoint beyond the up
    headroom: the top of the up request exceeds $P$ and is not delivered.
    (c) Attenuation of the down signal by $a_n^{-}$.}
    \label{fig:power}
\end{figure}

The operations layer acts on the cleared capacity, which is at most the bid;
we continue to write $\bvec{k}$ for it. It manages the SoC with two
mechanisms: the basepoint may exceed the headroom, up to a \emph{basepoint
limit}, and the regulation component of the request may be \emph{attenuated}
when it moves the SoC away from $\sstar$. Both lower the performance score. At
the first sample $jM$ of each basepoint period, the basepoint is the power that
returns the SoC to $\sstar$ in one period, clipped to the basepoint limit,
\begin{align}
    p_j &= \clip{\;\kappa\!\left(s_{jM}\right)
                   \left(s_{jM} - \sstar\right)\;}
                   {-\ell^{-}_j}{\;\ell^{+}_j},
    \label{eq:basepoint_policy}\\
    \kappa(s) &=
    \begin{cases}
        \eta_d\,E/\Tbp & s \ge \sstar,\\
        E/(\eta_c\,\Tbp) & s < \sstar,
    \end{cases}
    \nonumber
\end{align}
with $\ell_j^{\pm} \in [h_k^{\pm}, P]$, and the delivered power is
\begin{align}
    z_n = \Pi_{\mcZ(s_n,t_s)}\Big(&p_j
        + (1 - a_n^{+})\bidup{k} P\, \regup{n-\delta} \nonumber\\
        &- (1 - a_n^{-})\biddown{k} P\, \regdn{n-\delta}\Big),
    \label{eq:attenuation}
\end{align}
with attenuation factors $a_n^{\pm} \in [0,1]$, as shown in Fig.~\ref{fig:power}.

The \emph{headroom basepoint}, $\ell_j^{\pm} = h_k^{\pm}$ with $a_n^{\pm} = 0$, tracks the request with the
basepoint within the headroom and achieves a perfect score. Deviating from it
trades score for degradation; in what follows, we choose $\ell_j^{\pm}$ and
$a_n^{\pm}$. Both deviations move the SoC toward $\sstar$, so
Assumption~\ref{ass:nonadversarial} still holds and the predicted extremes of
Section~\ref{sec:bidding} remain valid. We analyze the up
market, in which the basepoint discharges the BESS when $s_{jM} > \sstar$;
analogous results hold for the down market, with $s_{jM} < \sstar$ and
$\eta_c$ in place of $\eta_d$. First, we derive bounds on $\ell_j^{\pm}$ and
$a_n^{\pm}$ given a worst-case score.

\begin{lemma}[Worst-case score bounds]\label{lem:deviation}
    Let $\bidup{k}, \biddown{k} > 0$, let the SoC bounds in~\eqref{eq:feasible}
    be inactive over basepoint period $j$, and let $a_n^{+} = 0$ whenever
    $p_j > h_k^{+}$ and $a_n^{-} = 0$ whenever $p_j < -h_k^{-}$. Given
    worst-case scores $\underline\sigma^{\pm} \in (\sigma_{\min}, 1]$, if
    \begin{align}
        \ell_j^{\pm} \le h_k^{\pm} + (1 - \underline\sigma^{\pm})\, b_k^{\pm} P,
        \qquad
        a_n^{\pm} \le 1 - \underline\sigma^{\pm},
        \label{eq:budget}
    \end{align}
    then $\sigma_n^{\pm} \ge \underline\sigma^{\pm}$ at every active sample of
    the period, and the qualification constraint~\eqref{eq:histscore} holds.
    Moreover, of the two mechanisms, $\sigma_n^{+}$ depends only on
    $\ell_j^{+}$ and $a_n^{+}$, and $\sigma_n^{-}$ only on $\ell_j^{-}$ and
    $a_n^{-}$.
\end{lemma}
\begin{proof}
    At a sample at which the up market is active, $\regdn{n-\delta} = 0$, and
    $z_n$ is $p_j + (1 - a_n^{+})\bidup{k}P\regup{n-\delta}$ clipped to
    $[-P, P]$, which is at least $p_j \ge -\ell_j^{-} \ge -P$. If
    $p_j \le h_k^{+}$, it is at most $h_k^{+} + \bidup{k}P = P$, so
    $|z_n - u_n| = a_n^{+}\bidup{k}P\regup{n-\delta}$. If $p_j > h_k^{+}$,
    then $a_n^{+} = 0$ and
    $|z_n - u_n| = [u_n - P]_+
    = [p_j - h_k^{+} - \bidup{k}P(1 - \regup{n-\delta})]_+$. In both cases
    $|z_n - u_n| \le (1 - \worstscoreup)\bidup{k}P$ by~\eqref{eq:budget}, and
    \eqref{eq:score_up} gives $\sigma_n^{+} \ge \worstscoreup$. The historical scores~\eqref{eq:histscore_def} are averages
    of scores at active samples, hence at least
    $\underline\sigma^{\pm} > \sigma_{\min}$.
\end{proof}
At every active sample, we satisfy the score bounds, as shown in
Lemma~\ref{lem:deviation}. Since
the revenue~\eqref{eq:revenue} and the qualification
constraint~\eqref{eq:histscore} depend on the scores only through averages, it
suffices to enforce them in expectation,
\begin{align}
    \E\big[\sigma_n^{\pm} \,\big|\, r^{\pm}_{n-\delta} > 0\big]
        \ge \underline\sigma^{\pm},
    \label{eq:expected_score}
\end{align}
at every sample, which admits looser bounds on $\ell_j^{\pm}$ and $a_n^{\pm}$.
This requires the distribution of the signal, which we estimate from
historical signals.
\begin{assumption}[Regulation signal]\label{ass:signal}
    Over each basepoint period, the distribution of $r^{\pm}_{n-\delta}$ is
    the same at every sample.
\end{assumption}
\begin{prop}[Expected-score bounds]\label{prop:admissible}
    Let Assumption~\ref{ass:signal} and the hypotheses of
    Lemma~\ref{lem:deviation} hold, with the scores $\underline\sigma^{\pm}$
    now required only in expectation. Let
    $\bar r^{\pm} = \E[r^{\pm}_{n-\delta} \mid r^{\pm}_{n-\delta} > 0]$ be the
    mean signal when the market is active, and let the \emph{cutoff} be
    \begin{align}
        r^{\pm}_{\mathrm{cut}} = \min\big\{c \in [0,1] :\ &
            \E\big[[r^{\pm}_{n-\delta} - c]_+ \,\big|\, r^{\pm}_{n-\delta} > 0\big]
            \nonumber\\
            &\le 1 - \underline\sigma^{\pm}\big\}.
        \label{eq:cutoff}
    \end{align}
    If
    \begin{align}
        \ell_j^{\pm} \le P - r^{\pm}_{\mathrm{cut}}\, b_k^{\pm} P,
        \qquad
        a_n^{\pm} \le \frac{1 - \underline\sigma^{\pm}}{\bar r^{\pm}},
        \label{eq:budget_expected}
    \end{align}
    then~\eqref{eq:expected_score} holds at every sample of the period.
    Moreover, $r^{\pm}_{\mathrm{cut}} \le \underline\sigma^{\pm}$ and
    $\bar r^{\pm} \le 1$, so the bounds~\eqref{eq:budget_expected} are at
    least the bounds~\eqref{eq:budget}, with equality if
    $r^{\pm}_{n-\delta} = 1$ whenever the market is active.
\end{prop}
\begin{proof}
    By the proof of Lemma~\ref{lem:deviation}, at a sample at which the up
    market is active, $|z_n - u_n|$ is at most
    $\bidup{k}P\,[\regup{n-\delta} - r^{+}_{\mathrm{cut}}]_+$ if $p_j > h_k^{+}$,
    and $a_n^{+}\bidup{k}P\,\regup{n-\delta}$ otherwise; both are at most
    $\bidup{k}P$, so $\sigma_n^{+} = 1 - |z_n - u_n|/(\bidup{k}P)$. Its
    conditional expectation is at least
    $1 - \E\big[[\regup{n-\delta} - r^{+}_{\mathrm{cut}}]_+ \mid \regup{n-\delta} > 0\big]$
    or $1 - a_n^{+}\bar r^{+}$, both at least $\worstscoreup$. The expected
    loss in~\eqref{eq:cutoff} is continuous and nonincreasing in $c$ and
    vanishes at $c = 1$, so $r^{+}_{\mathrm{cut}}$ is well defined. At
    $c = \worstscoreup$, the loss is at most $1 - \worstscoreup$ because
    $\regup{n-\delta} \le 1$, so $r^{+}_{\mathrm{cut}} \le \worstscoreup$, and
    $\bar r^{+} \le 1$, with equality if $\regup{n-\delta} = 1$ whenever the up
    market is active.
\end{proof}

\begin{table}[t]
\caption{Largest basepoint limit and attenuation factor that achieve the score $\underline\sigma^{\pm}$}\label{tab:admissible}
\centering
\begin{tabular}{lcc}
    & basepoint limit $\ell_j^{\pm}$ & attenuation $a_n^{\pm}$ \\
    \hline
    worst case & $P - \underline\sigma^{\pm} b_k^{\pm} P$
        & $1 - \underline\sigma^{\pm}$ \\
    expectation & $P - r^{\pm}_{\mathrm{cut}} b_k^{\pm} P$ & $(1 - \underline\sigma^{\pm})/\bar r^{\pm}$
\end{tabular}
\end{table}
The constraints on the limit and the attenuation factor are summarized in
Table~\ref{tab:admissible}.

Within these constraints, the remaining choice trades score for degradation.
\begin{assumption}[Operating degradation]\label{ass:running_degradation}
    Let $\map{\Dop}{[0,1]}{\realnonnegative}$ be differentiable, convex,
    nonincreasing on $[0, \sstar]$, and nondecreasing on $[\sstar, 1]$. The
    degradation incurred at sample $n$, in units of revenue, is $\Dop(s_n)$.
\end{assumption}
Given $s_{jM}$, the
basepoint limit and the attenuation factors of basepoint period $j$ solve
\begin{align}
    \max_{\ell_j^{\pm},\, a_n^{\pm}} \quad
    & \E\Big[\sum_{n=jM}^{(j+1)M-1} \big(R_n - \Dop(s_{n+1})\big)\Big]
    \label{eq:ops_objective}\\
    \text{s.t.}\quad
    & s_{n+1}=\FBess{s_n}{z_n},\quad
      \eqref{eq:basepoint_policy}\text{--}\eqref{eq:attenuation}, \nonumber\\
    & \ell_j^{\pm},\ a_n^{\pm} \text{ within Table~\ref{tab:admissible}}.
      \nonumber
\end{align}

We parametrize the basepoint limit by a function of the SoC,
$\map{f}{[0,1]}{[1/2, 1]}$, which we call the \emph{effective bid}, and set
\begin{align}
    \ell_j^{\pm} = \max\big\{h_k^{\pm},\ (1 - f(s_{jM}))P\big\}.
    \label{eq:ramp}
\end{align}
We call the basepoint~\eqref{eq:basepoint_policy} with this limit the
\emph{extended basepoint}. In real time, the BESS then behaves as if its bid
were $f(s_{jM})$: it keeps
$f(s_{jM})P$ for the signal and uses the rest of its rated power to restore
the SoC. Where $f \ge b_k^{\pm}$, the limit is the headroom, as in the
headroom basepoint, so $f = 1$ recovers it, and since $f \ge 1/2$, bids of
at most $1/2$ are unaffected. Where $f(s_{jM}) < \bidup{k}$ and the basepoint is at its limit,
the request exceeds $P$ exactly when $\regup{n-\delta} > f(s_{jM})/\bidup{k}$,
and by the proof of Lemma~\ref{lem:deviation},
\begin{align}
    |z_n - u_n| = \big[\bidup{k}P\,\regup{n-\delta} - f(s_{jM})P\big]_+,
    \label{eq:ramp_shortfall}
\end{align}
so the worst-case score is $f(s_{jM})/\bidup{k}$. By~\eqref{eq:ramp}, the
limit lies in the expectation row of Table~\ref{tab:admissible} if and only if
$f(s_{jM}) \ge \bidup{k} r^{+}_{\mathrm{cut}}$. The rest of this section
concerns the choice of $f$.
\begin{remark}[Tradeoff versus bidding]\label{rem:worst_case}
In the worst case, an effective bid below the bid is equivalent to bidding
less: at $\regup{n-\delta} = 1$, by~\eqref{eq:ramp_shortfall}, the up market
earns $f(s_{jM})P\pi_n^{+}$ at that sample, as a bid of $f(s_{jM})$ with a
perfect score would. Since the signal rarely reaches one, the effective bid
costs less in expectation.
\end{remark}
By Assumption~\ref{ass:signal}, the signal has the same distribution at every
sample of the basepoint period. In the probabilities below, inactive samples
count as $\regup{n-\delta} = 0$. Let
$\pi = \effpriceup/N$ be the mean price per sample.
\begin{prop}[Optimal effective bid]\label{prop:excess}
    Let Assumptions~\ref{ass:signal} and~\ref{ass:running_degradation} hold, let
    $s_{jM} > \sstar$, and replace $\Dop$ by its first-order approximation at
    $s_{jM}$ over the SoCs reached in basepoint period $j$. Over the period, the
    expected revenue minus the degradation is maximized by the
    limit~\eqref{eq:ramp} with $f(s_{jM}) = \bidup{k} c^\star$, where
    $c^\star$ is the largest
    $c \in [\max\{r^{+}_{\mathrm{cut}}, 1/(2\bidup{k})\}, 1]$ with
    \begin{align}
        \frac{\Prob(\regup{n-\delta} \ge c)}{\Prob(\regup{n-\delta} < c)}
            \ge \frac{\Tbp + t_s}{2}\,\frac{|\Dop'(s_{jM})|}{\eta_d E\, \pi},
        \label{eq:optimal_excess}
    \end{align}
    and $c^\star = \max\{r^{+}_{\mathrm{cut}}, 1/(2\bidup{k})\}$ if there is
    none. The optimal effective bid
    is nonincreasing in $[\sstar, 1]$.
\end{prop}
\begin{proof}
    Lowering $f$ by one unit lowers the signal level $f/\bidup{k}$ above which
    the request is not delivered. By~\eqref{eq:ramp_shortfall},
    it loses $P\pi\,\Prob(\regup{n-\delta} \ge f/\bidup{k})$ of revenue at
    each of the $M$ samples of the period. It also raises the limit by $P$,
    which in expectation discharges $P\,\Prob(\regup{n-\delta} < f/\bidup{k})$
    more at each sample and lowers the SoC by $t_s/(\eta_d E)$ per unit at that
    sample and every later one, so the degradation falls by
    $\frac{M(M+1)}{2}\frac{t_s P}{\eta_d E}|\Dop'(s_{jM})|
    \Prob(\regup{n-\delta} < f/\bidup{k})$. Comparing the two, with
    $(M+1)t_s = \Tbp + t_s$, lowering $f$ pays exactly
    when~\eqref{eq:optimal_excess} fails at $c = f/\bidup{k}$. The loss is
    convex and the saving concave in $f$, so the largest $c$ that
    satisfies~\eqref{eq:optimal_excess} is optimal. Since the left side
    of~\eqref{eq:optimal_excess} is nonincreasing in $c$ and, $\Dop$ being
    convex, the right side is nondecreasing on $[\sstar, 1]$, $c^\star$ is
    nonincreasing on $[\sstar, 1]$.
\end{proof}
The left side of~\eqref{eq:optimal_excess} is the odds that the signal
reaches $c$.
\begin{remark}[Degradation implied by an effective bid]\label{rem:implied_W}
Conversely, an effective bid $f$ is optimal for any degradation whose slope
satisfies, where $f$ decreases,
\begin{align*}
    |\Dop'(s)| = \frac{2\eta_d E \pi}{\Tbp + t_s}\,
        \frac{\Prob(\regup{n-\delta} \ge f(s)/\bidup{k})}{\Prob(\regup{n-\delta} < f(s)/\bidup{k})}.
\end{align*}
Since this slope is proportional to the price, for a given $\Dop$, a fixed
effective bid is optimal at only one price.
\end{remark}
\begin{remark}[Approximation]\label{rem:approx_W}
As in Remark~\ref{rem:overshoot}, replacing $\Dop$ by its first-order
approximation over a basepoint period introduces an error of order
$\Tbp/(E/P)$. The proposition also ignores the effect of the limit on later
basepoint periods.
\end{remark}

Computing the optimal effective bid requires $\Dop$ and the distribution of the
signal. Instead, we choose a piecewise linear $f$ that is nonincreasing in
$|s - \sstar|$, as in Proposition~\ref{prop:excess} and like the bids of
Fig.~\ref{fig:penalty}b, and use it in Algorithm~\ref{alg:ramp}.

The second mechanism attenuates the signal that moves the SoC away from
$\sstar$ by the factor $a_n^{\pm}$ in~\eqref{eq:attenuation}, as shown in
Fig.~\ref{fig:power}c. In the up market, it acts when $s_n < \sstar$,
where the up signal discharges the BESS: at an active sample,
$a_n^{+}\bidup{k}P\regup{n-\delta}$ is not delivered. By
Lemma~\ref{lem:deviation}, attenuating a signal spends the score of its
market, whereas the basepoint spends the score of the other market, so the two
mechanisms can be used at once.
\begin{prop}[Optimal attenuation]\label{prop:attenuation}
    Let Assumption~\ref{ass:running_degradation} hold, let the up market be
    active at sample $n$ with $s_n < \sstar$, and replace $\Dop$ by its
    first-order approximation at $s_n$. At sample $n$, the revenue minus the
    degradation is maximized by the largest $a_n^{+}$ of
    Table~\ref{tab:admissible} if
    \begin{align}
        \frac{t_s\,|\Dop'(s_n)|}{\eta_d E\,\pi_n^{+}} > 1,
        \label{eq:attenuation_threshold}
    \end{align}
    and by $a_n^{+} = 0$ otherwise.
\end{prop}
\begin{proof}
    By~\eqref{eq:score_up} and~\eqref{eq:revenue}, attenuating by $a_n^{+}$
    lowers the revenue by $a_n^{+}\bidup{k}P\regup{n-\delta}\pi_n^{+}$, and it
    raises the SoC by $t_s a_n^{+}\bidup{k}P\regup{n-\delta}/(\eta_d E)$, which
    lowers the degradation by $|\Dop'(s_n)|$ times this amount. Both are
    proportional to $a_n^{+}\bidup{k}P\regup{n-\delta}$, so the revenue minus
    the degradation is linear in $a_n^{+}$, and is maximized at the largest
    admissible $a_n^{+}$ if~\eqref{eq:attenuation_threshold} holds, and at zero
    otherwise.
\end{proof}
The threshold~\eqref{eq:attenuation_threshold} involves the SoC and the price
but not the signal, so we attenuate by a fraction that grows in steps with
$|s_n - \sstar|$, within the limit of Table~\ref{tab:admissible}, as in
Algorithm~\ref{alg:ramp}.
\begin{remark}[Order of the mechanisms]\label{cor:order}
Exceeding the headroom costs revenue only when the signal exceeds
$f(s_{jM})/\bidup{k}$,
whereas attenuation costs revenue at every active sample, so the basepoint is
used first and attenuation only far from $\sstar$.
\end{remark}
\begin{remark}[Amplification]\label{rem:amplification}
The signal that moves the SoC toward $\sstar$ could instead be amplified. By
the argument of Proposition~\ref{prop:attenuation}, this pays
when~\eqref{eq:attenuation_threshold} holds, but it spends the same score as
the basepoint, which restores the SoC at a lower cost, so we do not use it.
\end{remark}
\begin{algorithm}[t]
\caption{Extended basepoint with attenuation bands, for basepoint
period $j$}\label{alg:ramp}
\begin{algorithmic}[1]
\REQUIRE SoC $s_{jM}$; cleared bid $\bvec{k}$; effective bid $f$; attenuation
    bands
\STATE Set the basepoint limits $\ell_j^{\pm}$ by~\eqref{eq:ramp}
\STATE Compute $p_j$ by~\eqref{eq:basepoint_policy}
\FOR{each sample $n$ of the period}
    \STATE Set the attenuation factor of the signal that moves the SoC away
        from $\sstar$ from the band of $s_n$, and the other one to zero
    \STATE Deliver $z_n$ by~\eqref{eq:attenuation}
\ENDFOR
\end{algorithmic}
\end{algorithm}

\section{Numerical Results}\label{sec:numerics}
We first validate the predicted extremes (Section~\ref{sec:fit}), then
compare three ways to manage the SoC: bidding less
(Sections~\ref{sec:feasibility} and~\ref{sec:adaptive_bidding}), extending
the basepoint (Section~\ref{sec:effective_bid}), and a combination of both
(Section~\ref{sec:together}).

\subsection{Setup}
All experiments use a BESS with rated power $P = 2$\,MW, energy capacity
$E = 4$\,MWh, and efficiencies $\eta_c = \eta_d = 0.95$, with target SoC
$\sstar = 0.5$, retired when its state of health falls to $80\%$. The regulation signal is the PJM RegD signal of 2023, sampled
every $t_s = 2$\,s. This signal predates the split-market redesign, in which
up and down regulation are cleared separately; we take $\regup{n}$ and
$\regdn{n}$ as its positive and negative parts, which is the information the
split market provides at each sample. Since the 2023 market cleared a single
regulation price, we pay both markets at the same prices,
$\pi_{\mathrm{perf},n}^{+} = \pi_{\mathrm{perf},n}^{-}$ and
$\pi_{\mathrm{mil},n}^{+} = \pi_{\mathrm{mil},n}^{-}$; the effective prices
$\effpriceup$ and $\effpricedn$ in~\eqref{eq:effective_price} then differ only
through the time each market is active and its mileage. The SoC limits are
$s_{\min} = 0$ and $s_{\max} = 1$. Bid periods last $\Tbid = 30$\,min
($N = 900$), bids are committed $\Tbidadv = 35$\,min ahead, basepoint
periods last $\Tbp = 5$\,min ($M = 150$), and the response delay is
$\delta = 5$ samples ($10$\,s). A resource remains qualified with historical
scores of at least $\sigma_{\min} = 0.75$ in each market, and the risk level
is $\alpha = 0.05$. The historical score of each market is the rolling mean
over the preceding $100$\,h: the mean, over the last $200$ bid
periods, of the mean score over the active samples of each period. This
differs from~\eqref{eq:histscore_def} only in weighting each bid period
equally; it is still an average of scores at active samples.

Settlement uses the 2023 PJM regulation clearing prices from Data Miner~2,
at $5$-minute resolution, with the capability clearing price as
$\pi_{\mathrm{perf}}$ and the performance clearing price as $\pi_{\mathrm{mil}}$;
the portal has no prices for one day in August, which we ignore for revenue purposes.

The closed-loop runs of Sections~\ref{sec:feasibility}--\ref{sec:together}
simulate all of 2023. For each run, we report the revenue over the year, the mean scores
$\sigma^{\pm}$ over the samples at which each market is active, the battery
life, and the \emph{days at extreme SoC}, on which the SoC lies outside
$[0.4, 0.6]$, an interval centered at $\sstar = 0.5$. The battery life is the time until retirement. The state of health falls by
the sum, over the cycles of the realized SoC trajectory found by rainflow
counting, of the stress
\begin{align*}
    \frac{10^{-5}\, w}{1.4\, r^{-1/2} - 1.23}\, \exp\big(1.04\, |\bar s - \sstar|\big)
\end{align*}
of a cycle of depth $r$, mean SoC $\bar s$, and weight $w$, one for a full
cycle and $1/2$ for a half cycle~\cite[Eq.~(2)]{ZL-XW-KY-FZ:24}.
This model is independent of the degradation models of
Algorithm~\ref{alg:bidding}. It penalizes cycles at extreme SoC, but not the
time spent there, which accelerates calendar aging; the days at extreme SoC
measure it.

\subsection{Predicted extremes}\label{sec:fit}
We split the days of 2023 at random into training and test sets, with a $75/25$
split, so
that both span all seasons of the signal. From
each day of each set, we take $400$ overlapping bid periods of $30$\,min. We fit
the expected starting SoC~\eqref{eq:entry_estimate} and the affine
extremes~\eqref{eq:affine_split} on the training bid periods and evaluate the
extremes on the test bid periods.

\begin{figure}[htb]
    \centering
    \includegraphics[width=0.85\columnwidth]{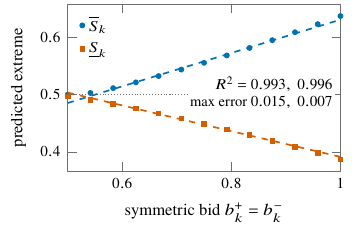}
    \caption{Predicted extremes~\eqref{eq:extremes} against a symmetric bid,
    $\alpha = 0.05$, with the least-squares affine
    fit~\eqref{eq:affine_split} over bids in $[1/2, 1]$ (dashed).}
    \label{fig:affine}
\end{figure}

The planes~\eqref{eq:affine_split} fitted on the training days are listed in
Table~\ref{tab:fit} and shown for symmetric bids in Fig.~\ref{fig:affine}. On the test bid periods, the planes match the extremes~\eqref{eq:extremes}
computed from the test signals to within $0.010$ for the lower extreme and
$0.015$ for the upper extreme. The SoC under~\eqref{eq:recovered} passes the
predicted lower and upper extremes in at most $2\%$ and $3\%$ of the test bid
periods, respectively, for every bid in $[1/2, 1]^2$, below $\alpha = 5\%$, as
required by Corollary~\ref{cor:limits}. As anticipated in
Remark~\ref{rem:conditioning}, $A$ is poorly conditioned: an error of $0.01$
in the extremes changes $\bidup{k} + \biddown{k}$ by at most $0.04$, but
$\bidup{k} - \biddown{k}$ by up to $0.45$.
\begin{table}[t]
\caption{Planes~\eqref{eq:affine_split} fitted on the training days, and
their largest errors in SoC on the training and test days}\label{tab:fit}
\centering
\setlength{\tabcolsep}{4pt}
\begin{tabular}{lrrrccc}
    & $g^{0}$ & $g^{+}$ & $g^{-}$ & $R^2$ & train & test \\
    \hline
    $\Sdn$ & $-0.116$ & $0.107$ & $0.118$ & $0.996$ & $0.007$ & $0.010$ \\
    $\Sup$ & $-0.161$ & $0.164$ & $0.128$ & $0.993$ & $0.015$ & $0.015$
\end{tabular}
\end{table}

\subsection{Managing the SoC}\label{sec:feasibility}
Constant bids with the headroom basepoint bound what bidding alone can achieve,
as shown in Table~\ref{tab:feasibility}. A bid of one earns the most, $703$\,k\$, at the
largest cost to the battery: it lasts $2.83$ years, with $302$ days at extreme
SoC. A bid of $1/2$ saves the battery the most: its headroom absorbs nearly all
of the drift, as in Lemma~\ref{lem:absorbed}, the SoC never leaves $[0.4, 0.6]$, and
the battery lasts $5.73$ years, but it earns only $353$\,k\$.

Even a small headroom helps. A bid of $0.99$ instead of one gives up $0.7\%$
of the revenue, nearly halves the days at extreme SoC, from $302$ to $161$, and
lengthens the battery life from $2.83$ to $3.34$ years.

\begin{table}[t]
\caption{Constant bids with the headroom basepoint over 2023. Mean scores
over the active samples of each market; days at extreme SoC: days with the
SoC outside $[0.4, 0.6]$; battery life from rainflow counting.}\label{tab:feasibility}
\centering
\setlength{\tabcolsep}{4pt}
\begin{tabular}{lrccrr}
    & revenue & mean & mean & days at & battery \\
    bid & [k\$] & $\sigma^{+}$ & $\sigma^{-}$ & extreme SoC & life [y] \\
    \hline
    $1$ & $703$ & $1.00$ & $0.99$ & $302$ & $2.8$ \\
    $0.99$ & $698$ & $1.00$ & $1.00$ & $161$ & $3.3$ \\
    $1/2$ & $353$ & $1.00$ & $1.00$ & $0$ & $5.7$
\end{tabular}
\end{table}

\subsection{Adaptive bidding}\label{sec:adaptive_bidding}
We bid as per Algorithm~\ref{alg:bidding}, which decides how much less to bid,
one bid period at a time. It sets $\effpriceup = \effpricedn = \hat\pi$, the
\emph{expected price}, in \$/MW, which it assumes when committing a bid;
settlement uses the realized 2023 prices. Since the expected price is a forecast, we report
the results for $\hat\pi \in \{15, 20, 25, 30, 35\}$\,\$/MW.

Since the regulation signal largely determines the cycling, we penalize
instead how far the SoC strays from $\sstar$. We consider two styles of
degradation model: one penalizes only the extreme farther from $\sstar$, as
in~\eqref{eq:penalty}, and the other penalizes both extremes. As instances, we
consider two models which we refer to as the \emph{exponential} and the \emph{quartic} models respectively,
\begin{align*}
    \Dbid(s_1, s_2) &= \lambda\, \frac{\exp(8d) - 1}{\exp(4) - 1},
    \; d = [\max\{\sstar - s_1,\ s_2 - \sstar\}]_+,\\
    \Dbid(s_1, s_2) &= \lambda\, \big([\sstar - s_1]_+^4 + [s_2 - \sstar]_+^4\big).
\end{align*} For each model, $\lambda$ is chosen so that
at $\hat\pi = 25$\,\$/MW the bid is full for $\hat s_k \in [0.4, 0.6]$.
We run symmetric and split bids, capped at $0.99$ unless stated otherwise, with
the headroom basepoint.

For the exponential model, the bids have closed forms, derived in
Appendix~\ref{app:closed_forms}; for the quartic model, we find them by
bisection.

We draw four findings from these runs.

\begin{figure*}[t]
    \centering
    \includegraphics[width=\textwidth]{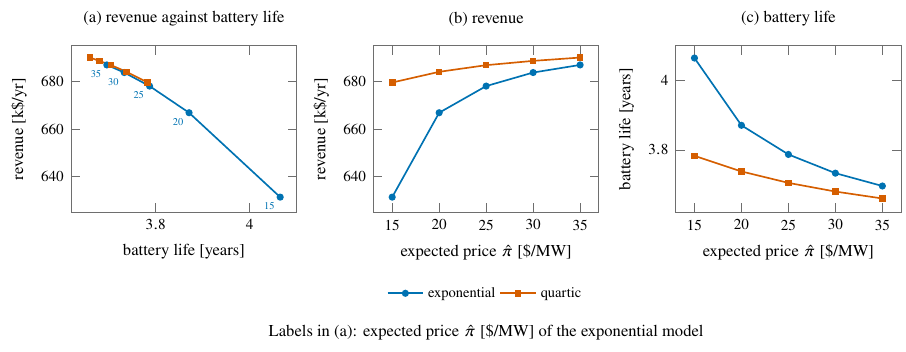}
    \caption{The expected price as a knob: symmetric bids of
    Algorithm~\ref{alg:bidding} over 2023, capped at $0.99$, with the headroom
    basepoint. (a) Revenue against battery life; (b) revenue and (c) battery
    life against the expected price $\hat\pi$ assumed by the bidding layer.
    Settlement uses the realized prices.}
    \label{fig:results_bidding}
\end{figure*}

\paragraph{The expected price controls the revenue--aging trade-off}
With symmetric bids and the exponential model, raising $\hat\pi$ from $15$ to
$35$\,\$/MW raises the revenue from $631$ to $687$\,k\$ per year, shortens the
battery life from $4.07$ to $3.70$ years, and raises the days at extreme SoC
from $12$ to $56$, as shown in Fig.~\ref{fig:results_bidding}. Against the constant bid of $0.99$
of Table~\ref{tab:feasibility}, even the run at $35$\,\$/MW gives up only
$1.5\%$ of the revenue, $687$ against $698$\,k\$, for $11\%$ more battery
life, $3.70$ against $3.34$ years, and about a third of the days at extreme
SoC, $56$ against $161$.

\paragraph{Degradation model}
The two models are comparable. At a given expected price, the quartic model
bids more: $28$ against $12$ days at extreme SoC at $\hat\pi = 15$\,\$/MW. At
equal revenue, it spends slightly fewer days at extreme SoC, $38$ against $44$
at $684$\,k\$. Both models stay off the SoC limits at every expected price. We use the exponential model in the rest of this section, since its
bid is a closed form, piecewise linear in $\hat s_k$~\eqref{eq:exp_bid},
whereas the quartic bid requires bisection.

\begin{figure}[htb]
    \centering
    \includegraphics[width=\columnwidth]{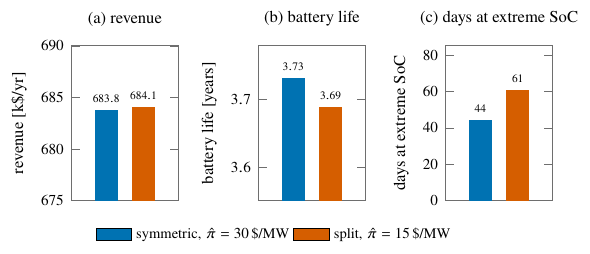}
    \caption{Symmetric against split bids at equal revenue, $684$\,k\$, with
    the exponential model, bids capped at $0.99$, and the headroom basepoint:
    (a) revenue, (b) battery life, and (c) days at extreme SoC. The axes of
    (a) and (b) do not start at zero.}
    \label{fig:split}
\end{figure}

\paragraph{Symmetric versus split bids}
Split bids hold the direction that restores the SoC at one. We compare them
with symmetric bids at equal revenue: symmetric bids at
$\hat\pi = 30$\,\$/MW against split bids at $15$\,\$/MW, both at about
$684$\,k\$. At the same revenue, split bids shorten the battery life by
$1.2\%$ and add $16$ days at extreme SoC, as shown in Fig.~\ref{fig:split}. With equal up
and down prices, as in 2023, splitting the bids buys nearly nothing, about
$0.3$\,k\$ per year.

\begin{figure}[htb]
    \centering
    \includegraphics[width=\columnwidth]{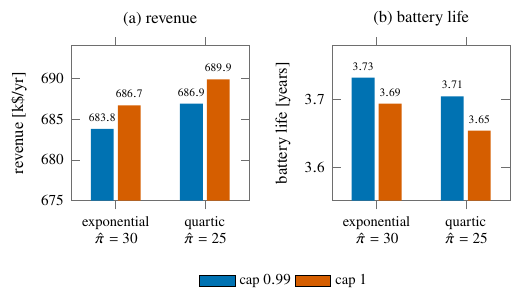}
    \caption{Symmetric bids capped at $0.99$ and at one, with the headroom
    basepoint, for representative runs of the two models: (a) revenue and
    (b) battery life. The axes do not start at zero.}
    \label{fig:cap}
\end{figure}

\paragraph{Leaving some headroom}
As for the constant bids of Section~\ref{sec:feasibility}, a small headroom
pays off. To compare the caps, we take one representative run of each model, aligned
at similar revenue: the exponential model at $\hat\pi = 30$\,\$/MW and the
quartic model at $\hat\pi = 25$\,\$/MW. Capping the symmetric bids at one
instead of $0.99$ raises the revenue by $0.4\%$ for both, but shortens the
battery life by $1.0\%$ and $1.4\%$, respectively, and raises the days at
extreme SoC by $29\%$ and $33\%$, as shown in Fig.~\ref{fig:cap}. The revenue rises by only $0.4\%$, not $1\%$, because
the cap binds only in bid periods in which Algorithm~\ref{alg:bidding} bids
full; in the others, it bids less, so the mean bid rises only from $97.2$ to
$97.6\%$ for the exponential model and from $97.6$ to $98.0\%$ for the quartic
model.

\subsection{Extended basepoint}\label{sec:effective_bid}
We now focus on the operations layer alone: we hold the bid constant and replace the headroom basepoint by the extended
basepoint of Algorithm~\ref{alg:ramp}, with the effective bid $f$ equal to one on $[0.4, 0.6]$, to $1/2$ outside
$[0.25, 0.75]$, and linear in between. The degradation model for
which this $f$ is optimal is derived in Appendix~\ref{app:implied}. Unless stated otherwise, we do
not attenuate; when we do, the bands deliver $100\%$, $90\%$, and $70\%$ of the
signal that moves the SoC away from $\sstar$ on $[0.4, 0.6]$, on
$[0.3, 0.4] \cup [0.6, 0.7]$, and outside $[0.3, 0.7]$, respectively.

\begin{figure}[htb]
    \centering
    \includegraphics[width=\columnwidth]{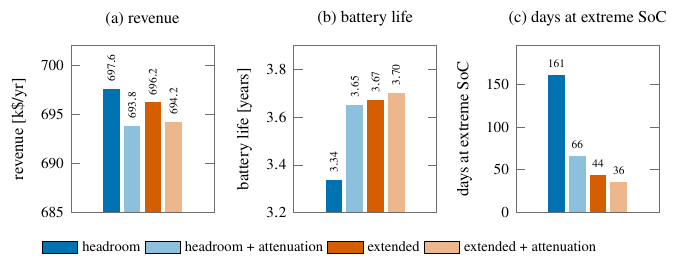}
    \caption{The operations layer at a constant bid of $0.99$: the headroom
    and the extended basepoint of Algorithm~\ref{alg:ramp}, each with and
    without attenuation. (a) Revenue, (b) battery life, and (c) days at
    extreme SoC. The axes of (a) and (b) do not start at zero.}
    \label{fig:effbid}
\end{figure}

The results are shown in Fig.~\ref{fig:effbid}, and we draw three findings
from them.

\paragraph{The extended basepoint lengthens the battery life at a small cost in revenue}
Replacing the headroom basepoint by the extended basepoint changes the revenue of
constant bids by at most $0.2\%$, from $697.6$ to $696.2$\,k\$ at a bid of
$0.99$ and from $702.7$ to $702.8$\,k\$ at a bid of one, and keeps both mean
scores at $0.996$ or above. It lengthens the battery life from $3.34$ to
$3.67$ years and from $2.83$ to $3.59$ years, and cuts the days at extreme SoC
from $161$ to $44$ and from $302$ to $76$.

\paragraph{Use the basepoint before attenuation}
With the headroom basepoint, attenuation also restores the SoC: at a bid of
$0.99$, it lengthens the battery life from $3.34$ to $3.65$ years and cuts the
days at extreme SoC from $161$ to $66$, for $0.5\%$ of the revenue. The
extended basepoint alone does better on all three counts, with $696.2$
against $693.8$\,k\$, $3.67$ against $3.65$ years, and $44$ against $66$ days,
as expected, and discussed in Remark~\ref{cor:order}. On top of the extended basepoint,
attenuation adds $0.03$ years of battery life and cuts $8$ days at extreme
SoC, for $0.3\%$ of the revenue, and keeps both mean scores at $0.99$ or
above.

\paragraph{Managing the SoC in real time beats bidding less}
A constant bid of $0.99$ with the extended basepoint spends $44$ days at extreme
SoC, as many as adaptive bidding at $\hat\pi = 30$\,\$/MW in
Section~\ref{sec:adaptive_bidding}, but earns $696$ against $684$\,k\$,
$1.8\%$ more, for a battery life of $3.67$ against $3.73$ years. As
expected, and discussed in Remark~\ref{rem:worst_case}, the extended basepoint loses revenue only
when the signal is large, whereas a lower bid loses it at every sample.

\subsection{Adaptive bidding and extended basepoint together}\label{sec:together}
We run Algorithm~\ref{alg:bidding} with the exponential model, symmetric bids
capped at $0.99$, and the extended basepoint, at the expected prices of
Section~\ref{sec:adaptive_bidding}.

We draw two findings from these runs.

\begin{figure}[htb]
    \centering
    \includegraphics[width=\columnwidth]{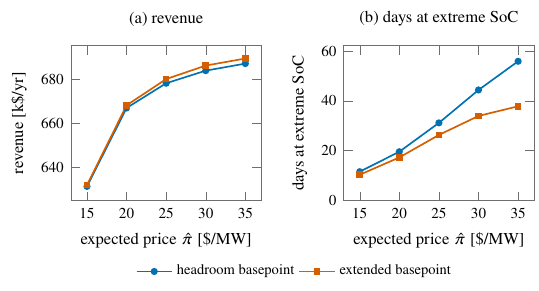}
    \caption{Adaptive bidding with the headroom and with the extended
    basepoint: symmetric bids, exponential model, capped at $0.99$. (a) Revenue
    and (b) days at extreme SoC against the expected price $\hat\pi$.}
    \label{fig:together}
\end{figure}

\paragraph{The extended basepoint improves adaptive bidding}
At every expected price, replacing the headroom basepoint by the extended
basepoint raises the revenue by $0.6$ to $2.3$\,k\$, changes the battery life
by at most $0.02$ years, and cuts the days at extreme SoC by $11$ to $32\%$,
as shown in Fig.~\ref{fig:together}.

\paragraph{The bidding layer buys battery life at a poor rate}
Against the extended basepoint alone at a constant bid of $0.99$, adaptive
bidding at $\hat\pi = 35$\,\$/MW gives up $1.0\%$ of the revenue, $689.3$
against $696.2$\,k\$, for $1.1\%$ more battery life, $3.72$ against
$3.67$ years, and $6$ fewer days at extreme SoC. The two layers are not
designed together: as in Assumption~\ref{ass:nonadversarial}, the bidding layer
predicts the extremes as if only the headroom restored the SoC, so it bids
low against excursions that the extended basepoint absorbs.

\section{Discussion}\label{sec:discussion}

\paragraph{The layers in tandem}
As shown in Section~\ref{sec:together}, the extended basepoint improves
adaptive bidding at every expected price, but the two layers do not divide the
work well. As in Assumption~\ref{ass:nonadversarial}, the bidding layer
predicts the extremes as if only the headroom restored the SoC, so it bids low against excursions
that the extended basepoint absorbs, and it buys battery life at a poor rate.
Layers designed in isolation compound their conservatism.

\paragraph{Feedforward and feedback}
The bid acts as feedforward: it is committed on a prediction of the extremes,
it scales the drift~\eqref{eq:drift} and sets the headroom available to the
basepoint, and it costs revenue at every sample. The basepoint acts as
feedback: it responds to the measured SoC and, when extended, costs revenue
only when the signal is large, as discussed in Remark~\ref{rem:worst_case}. On
the 2023 RegD signal, feedback is the cheaper lever, as shown in
Section~\ref{sec:effective_bid}. At
the best division of the work, the revenue lost per unit of SoC restored would
be equal for the two levers; condition~\eqref{eq:optimal_excess} gives this
cost for the extended basepoint.

\paragraph{Learning the extremes from closed-loop data}
The bidding layer sees the operations layer only through the
recovery~\eqref{eq:recovery}. Replacing the headroom by the basepoint limit
in~\eqref{eq:recovery} would account for the extended basepoint, but the
recovery would then depend on the SoC, and the convexity of
Proposition~\ref{prop:convex} would no longer follow. Instead, $\hat s_k$ and
the affine extremes~\eqref{eq:affine_split} can be fitted on SoC trajectories
simulated under the operations layer in use. The bid problem stays convex,
since the fit is affine, and the bidding layer adapts to the operations layer,
at the cost of the guarantee of Corollary~\ref{cor:limits}.

\section{Conclusion}
We studied how a BESS in the split PJM regulation market should manage its
SoC, with two levers: the bids, committed before the regulation signal is
known, and the basepoint, set in real time. The bidding layer bids against
conservatively predicted extreme SoCs, in a convex problem with closed-form
solutions, and its expected price trades revenue for battery life. The
operations layer extends the basepoint beyond the headroom at a small cost in
score. On the 2023 RegD signal, the extended basepoint lengthens the battery
life of constant bids by $10$ to $27\%$ at a change in revenue of at most
$0.2\%$, and earns more than adaptive bidding at equal time at extreme SoC:
the extended basepoint loses revenue only when the signal is large, whereas a
lower bid loses it at every sample.

We discuss in Section~\ref{sec:discussion} how to design the two layers together;
models of regulation prices would also replace the expected price by a
forecast.

\appendices
\section{Closed-form bids}\label{app:closed_forms}
For the exponential model of Section~\ref{sec:adaptive_bidding}, the bids of
Algorithm~\ref{alg:bidding} have closed forms. Since~\eqref{eq:interior_distance} does not involve $\hat s_k$,
the symmetric bid is piecewise linear in $\hat s_k$, as shown in
Fig.~\ref{fig:penalty}b:
\begin{align}
    \beta_k^\star &= \clip{\min\{\beta^{\uparrow}, \beta^{\downarrow}\}}{1/2}{1}, \; \text{where}
    \nonumber\\
    \beta^{\uparrow} &= \frac{\frac{1}{8}\ln\frac{\langle \vc, \vectorones[2]\rangle(\exp(4) - 1)}
        {8\lambda (g^{+}_{\uparrow} + g^{-}_{\uparrow})} - (\hat x_k^{\uparrow} - \sstar)}
        {g^{+}_{\uparrow} + g^{-}_{\uparrow}},
    \label{eq:exp_bid}\\
    \beta^{\downarrow} &= \frac{\frac{1}{8}\ln\frac{\langle \vc, \vectorones[2]\rangle(\exp(4) - 1)}
        {8\lambda (g^{+}_{\downarrow} + g^{-}_{\downarrow})} - (\sstar - \hat x_k^{\downarrow})}
        {g^{+}_{\downarrow} + g^{-}_{\downarrow}},
    \nonumber
\end{align}
further limited by the SoC limits of~\eqref{eq:symmetric_problem}. Here
$\beta^{\uparrow}$ and $\beta^{\downarrow}$ satisfy~\eqref{eq:interior_distance}
with the upper and the lower extreme the farther one, respectively; the minimum
is optimal unless both extremes are equally far at the optimum, and otherwise
errs toward a smaller bid. For split bids with $\hat s_k < \sstar$
and the lower extreme the farther one, the same argument places the lower
extreme at the distance
$d = \frac{1}{8}\ln\big(P\worstscoreup\effpriceup(\exp(4) - 1)/(8\lambda
g^{+}_{\downarrow})\big)$ from $\sstar$, so the bid is
\begin{align}
    \bidup{k} = \clip{\frac{d - (\sstar - \hat x_k^{\downarrow}) - g^{-}_{\downarrow}}
        {g^{+}_{\downarrow}}}{1/2}{1},
    \label{eq:one_sided}
\end{align}
further limited by the SoC limits of~\eqref{eq:simplified}, and symmetrically
when $\hat s_k \ge \sstar$. For the quartic model, both extremes contribute to
the right derivative of Corollary~\ref{cor:symmetric}, which is then a cubic
polynomial in the bid; we find the bid by bisection.

\section{Degradation implied by the chosen effective bid}\label{app:implied}
\begin{figure}[htb]
    \centering
    \includegraphics[width=\columnwidth]{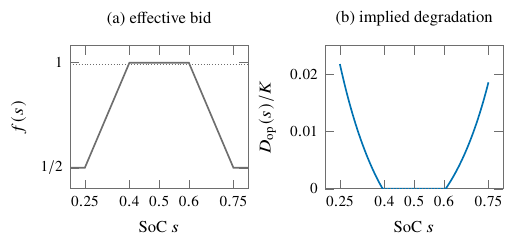}
    \caption{Degradation implied by the effective bid, for the 2023 RegD signal
    and a constant bid of $0.99$: (a) the effective bid $f$, with the bid
    dotted; (b) the implied $\Dop$, in units of $K = 2\eta E\pi/(\Tbp + t_s)$,
    from the edge of the full band.}
    \label{fig:implied}
\end{figure}

The effective bid $f$ of Section~\ref{sec:effective_bid} is tuned by hand. A
natural question is therefore which degradation model $\Dop$ makes this $f$
optimal. By Remark~\ref{rem:implied_W}, $f$ is optimal for any degradation
model whose slope, where $f$ decreases, is proportional to the odds that the
signal exceeds $f(s)/b$. We compute this slope for the 2023 RegD signal and a
constant bid $b = 0.99$ in three steps.
\begin{enumerate}
    \item Estimate the tail $q^{\pm}(c) = \Prob(r^{\pm}_n \ge c)$ of each
    market's signal over all samples of 2023, with inactive samples counting as
    zero.
    \item Where $f$ decreases, by Remark~\ref{rem:implied_W}, the slope is
    $|\Dop'(s)| = K\, q(c)/(1 - q(c))$, with $c = f(s)/b$ and
    $K = 2\eta E\pi/(\Tbp + t_s)$; the up market applies above $\sstar$, with
    $\eta = \eta_d$, and the down market below, with $\eta = \eta_c$.
    \item Integrate the slope from the edge of the full band $[0.4, 0.6]$.
\end{enumerate}
The result is shown in Fig.~\ref{fig:implied}. On $[0.4, 0.6]$, $f$ exceeds the
bid, so the basepoint stays within the headroom and the model is only bounded:
its slope is at most $K q(1)/(1 - q(1))$, and the signal reaches one in only
$2.6\%$ of the samples of either market. Along the ramps, the implied slope
rises from $0.07K$ to $0.21K$ above $\sstar$ and from $0.09K$ to $0.24K$ below,
so the implied model is convex, as Assumption~\ref{ass:running_degradation}
requires, and its marginal cost roughly triples across each ramp. Beyond $0.25$
and $0.75$, $f$ is constant and the model is not identified. The scale $K$
grows with the price, so a fixed effective bid is optimal at only one price.

The same signal statistics check the score bounds of
Section~\ref{sec:operations}. At $\underline\sigma = 0.75$, the cutoffs are
$r^{+}_{\mathrm{cut}} = 0.22$ and $r^{-}_{\mathrm{cut}} = 0.24$, and the mean
active signals are $\bar r^{+} = 0.42$ and $\bar r^{-} = 0.44$. The floor
$f \ge 1/2$ thus exceeds the bound $f \ge \bidup{k} r^{+}_{\mathrm{cut}}$
of Section~\ref{sec:operations}, and the strongest attenuation band, $30\%$, lies
within the limit $(1 - \underline\sigma)/\bar r^{\pm} \ge 0.57$ of
Table~\ref{tab:admissible}.

\bibliography{alias, Main, FB, New}

\end{document}